# Direct *in situ* determination of the surface area and structure of deposited metallic lithium within lithium metal batteries using ultra small and small angle neutron scattering

Christophe Didier[1], Elliot P. Gilbert[1], Jitendra Mata[1] and Vanessa Peterson[1]*

[1] *Australian Centre for Neutron Scattering, Australian Nuclear Science and Technology Organization, Locked Bag 2001, Kirrawee DC, NSW 2232, Australia*

## Abstract

Despite being the major cause of battery safety issues and detrimental performance, a comprehensive growth mechanism for metallic lithium deposited at electrode surfaces in lithium metal batteries remains elusive. While lithium surface morphology is often derived indirectly, here, detailed information is directly obtained using in situ small and ultra-small angle neutron scattering, in bulk and non-destructively. Features of 1-10 μm and 100-300 nm are identified; the latter contribute to most of the surface area and their size inversely correlates to applied current density. Surface area per unit volume increases continuously during charging from 1-4 h at 2 mA/cm$^2$ but more slowly during discharge. Comparatively higher values are reached after just 1 h at 20 mA/cm$^2$ which remain constant in subsequent cycles. Such quantitative insight into the processes of metallic lithium growth within batteries may enable the development of safer high performance lithium metal batteries.

## 1. Introduction

Considerable effort has been devoted to the improvement of lithium-ion batteries (LIBs) for the past 30 years,[1] enabling the use of portable electronics and electric vehicles. There is interest in replacing commonly used LIB electrodes such as graphite with lithium metal because of its order-of-magnitude larger specific capacity; however, rechargeable lithium metal batteries (LMBs) are often plagued by low efficiency and rapid capacity fade.[2,3] These LMB performance issues are attributed primarily to the formation of high surface area microstructures at the lithium surface, eventually creating short-circuits between electrodes or irreversibly separating from the electrode after partial dissolution resulting in electrochemically inactive "dead lithium".[2,4,5]

Many factors influence the morphology of deposited lithium in a LMB,[6] however despite considerable research, the mechanism of lithium deposition and microstructure development in LMBs is still poorly understood, partly because observing deposited lithium is experimentally challenging. Historically, the deposited lithium is examined *post mortem* after extraction from the LMB - a mechanical process potentially changing the electrode surface. *In situ* techniques, where lithium is examined within the LMB, have enabled remarkable progress in understanding the parameters that influence lithium growth, notably using optical and electron microscopies, [3,7-13] however those methods require model cells that may not accurately represent the chemical environment within typical LMBs.[14]

Electron and optical microscopy studies have revealed a range of deposited lithium morphologies, with the most commonly reported structures in LMBs with liquid electrolytes being so-called whiskers, mosses, and dendrites, while noting a lack of naming convention.[6,8-10,13,15] Whiskers are reported to initially appear as needles approximately 100 nm wide and up to 10 μm long. (Figure 1) The description of mossy lithium is given to a porous layer up to several hundreds of microns thick and comprising interconnected objects with diameter approximately 0.1 to 10 μm. Mossy lithium is reported to arise from the interweaving and broadening of whiskers,[7,10,13] however, it is unclear whether all whiskers become mossy lithium or if both coexist. Dendrites are reported as 100-500 μm long fractal-like filaments approximately 1 μm thick, sometimes forming dense bushes.[9]

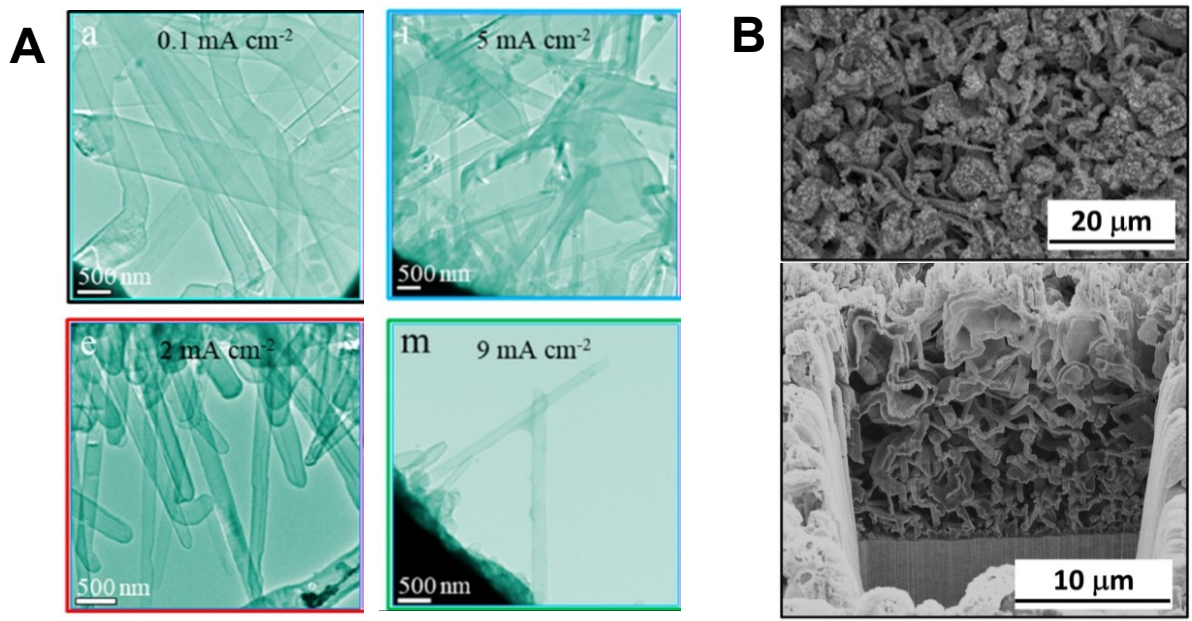

**Figure 1**: Representative microscopy images of A) Lithium whiskers observed using bright field cryo-transmission electron microscopy. Adapted with permission from Xu et al.[15] Copyright 2020 American Chemical Society. B) Surface (top) and cross-section obtained using a focussed ion beam (bottom) of mossy lithium interconnected structures at 0.1 MPa using cryo-scanning electron microscopy. Reprinted from Harrison et al.[16] copyright 2021, with permission from Elsevier.

Small- and ultra-small-angle neutron scattering (SANS and USANS) are techniques that can be used to study the morphology of structures within objects such as battery materials on length scales typically ranging from 1 to 10000 nm, falling within the range reported for deposited lithium structures. These techniques are sensitive to neutron scattering length density (SLD) inhomogeneities in a sample, with the scattered intensity proportional to the amount of inhomogeneities and contrast around them given by the square of the difference of the SLD. SLD inhomogeneities arise from elemental and isotopic density variations, as found at the interface between two phases. Scattering intensity variations with scattering vector $Q$ depend on the spatial distribution of such inhomogeneities, which can be related to the size and shape of those objects.

There are large advantages to studying LIB components including LMBs using SANS and USANS, where the high penetration of neutrons easily permits full transmission *in situ* measurements of components within typical electrochemical cells where information is averaged over the entire cell, in contrast to microscopy studies. Despite these advantages, only a limited number of *in situ* SANS studies of batteries have been performed, and none with USANS. Relatively good sensitivity to electrode surface changes have been observed in several cases using SANS, such as lithium sulfide deposition within porous carbon,[17] SEI formation at the surface of lithium titanate,[18] or interfaces between lithiated graphite phases.[19,20]

A symmetrical lithium metal pouch cell was studied using SANS[20] and an increase of the total integrated intensity after cycling reported, confirming the sensitivity of SANS to lithium electrode changes, but where details of the lithium morphology were not derived. *In situ* SANS from a custom lithium metal cell with solid-state electrolyte LLZNO also showed a small increase in scattering intensity and the formation of lithium features 1-10 nm in size.[21] Here, we assess the applicability of in situ SANS and USANS to study the morphology and growth process of metallic lithium deposited within a symmetrical LMB. We first characterise the signal from individual components to further guide construction of the cell that we use; we subsequently evaluate simple models to describe the *in situ* SANS and USANS data and derive for the first time parameters such as surface area and particle size of deposited lithium structures after applied galvanostatic cycling, over an electrode area that is representative of the whole cell.

## 2. Individual cell component scattering and cell construction

A flat laminated pouch cell construction, similar to that used in other work,[19,20] was chosen for its relatively simple and flexible assembly (Figure 2). Because the neutron beam passes through all cell components, all components may contribute to neutron scattering and attenuation. Therefore, battery components were measured individually on both the Kookaburra (USANS) and Quokka (SANS) instruments at ACNS, ANSTO[22,23] and selected for the construction of the *in situ* pouch cell based on the following criteria: isotropic scattering – since anisotropic scattering leads to irreproducible data in slit geometry instruments (e.g. Bonse-Hart type USANS instruments such as Kookaburra) - low coherent scattering by electrochemically inactive components, low neutron attenuation and negligible multiple scattering. Four current collectors were compared: nickel mesh, copper mesh, electrodeposited copper foil with one rough surface, and roll-annealed copper foil with smooth surfaces. Current collectors typically have high surface roughness on one side to promote electrode adherence.[24]

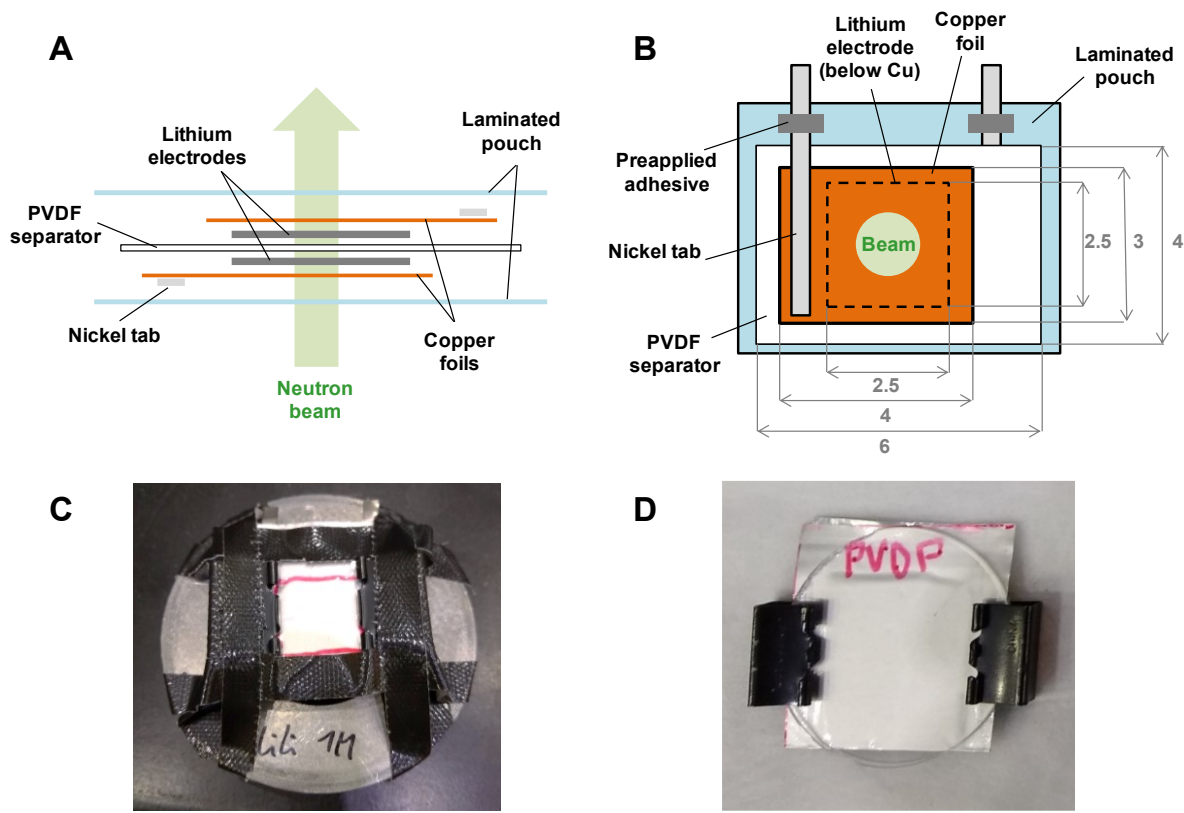

**Figure 2**: A) Side view schematic with exaggerated distances of the symmetrical lithium pouch cell. B) Front view schematic with the top laminated pouch omitted for clarity, distances are in cm. The 1.2 cm diameter neutron beam corresponds to that on the Kookaburra instrument. C) Photograph of the *in situ* cell covered with quartz slides held by bulldog clips (behind tape) mounted on the metallic sample holder. D) Photograph of the electrolyte-wet PVDF sealed in a laminated pouch between quartz slides held by bulldog clips.

Both metal mesh current collectors exhibited anisotropic and relatively intense coherent scattering (Figure S1), consistent with other metal mesh materials,[25] and were considered unsuitable for the *in situ* cell. Scattering per unit area (scattering per unit volume multiplied by the thickness) from rough electrodeposited foil was greater by an order of magnitude compared to the smooth roll-annealed foil, despite being a third of the thickness (Figure 3A); this behaviour presumably arises from surface scattering. SANS between scattering vector $Q$ of $10^{-3}$ and $10^{-1}$ Å$^{-1}$ follows a slope of $Q^{-3.838(15)}$ and $Q^{-3.548(18)}$ for 'rough' copper foil and for 'smooth' copper foil, respectively. A slope close to $Q^{-4}$ (Porod's law) suggests surface scattering as has previously been reported from scratched metal plates[26] as expected from the 'rough' foil. The slope close to $Q^{-3.5}$ for the 'smooth' foil may indicate a mixture of scattering arising from dislocations within the bulk where $Q^{-3}$ variation is expected, and scattering by surfaces, pores, or impurities with $Q^{-4}$ variation, as observed in bulk copper and other metals.[26-28] We note that the terms 'rough' and 'smooth' here refer to macroscopically observed characteristics of the foils; this is not to be confused with the concepts of smooth and rough surfaces at the nanoscale level which exhibit Porod-power law scattering with exponent -4 and greater than -4 (e.g. -3.5) respectively. Smooth foil was selected as the current collector for the *in situ* cell considering the relatively small and isotropic coherent scattering, with negligible neutron attenuation. (Table S1)

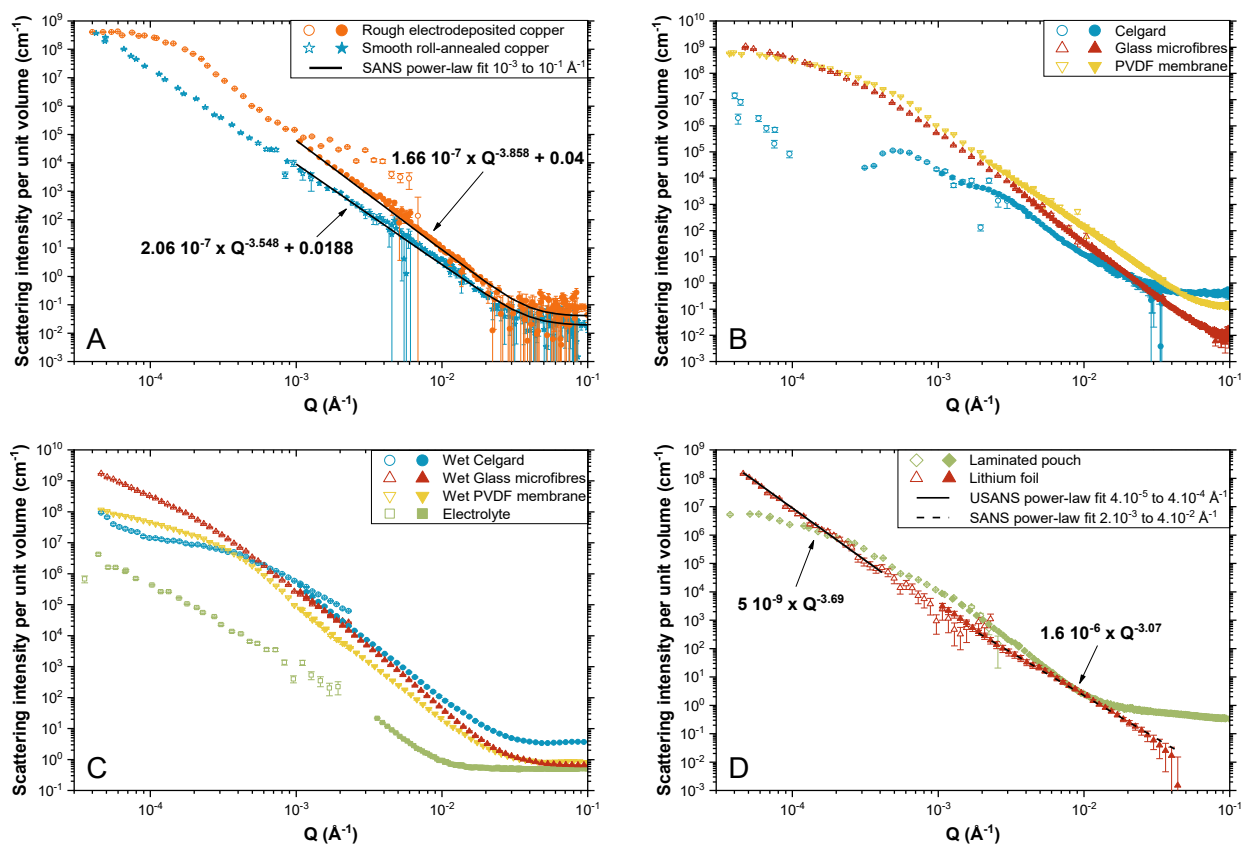

**Figure 3**: Desmeared USANS (open symbols) and SANS (closed symbols) differential scattering cross-section per unit volume of A) rough electrodeposited and smooth roll-annealed copper foils. Power-law exponents were extracted by fitting of the function $A \times Q^{-n} + B$ over the SANS data between $10^{-3} < Q < 10^{-1}$ Å$^{-1}$. Refined values were $A = 1.66(14) \times 10^{-7}$, $B = 0.04(6)$ and $n = 3.858(15)$ for rough foil and $A = 2.06(19) \times 10^{-7}$, $B = 0.0188(14)$ cm$^{-1}$ and $n = 3.548(18)$ for smooth foil. B) Celgard, glass microfibre, and polyvinylidene fluoride (PVDF) membranes measured in air, C) electrolyte-wet Celgard, glass microfibres, PVDF membrane, and the 1 M LiPF$_6$ in ethylene carbonate/dimethyl carbonate electrolyte, shown after subtraction of scattering from the laminated pouch, with anisotropic two-dimensional data for Celgard shown inset, and D) lithium metal after subtraction of scattering from the laminated pouch and laminated pouch data. Power-law exponents for lithium foil were extracted by fitting of the function $A \times Q^{-n}$ over SANS and smeared USANS data within the $Q$ range shown in legend. Refined values were $A = 5(2) \times 10^{-9}$ and $n = 3.69(4)$ for USANS and $A = 1.6(4) \times 10^{-6}$ and $n = 3.07(5)$ for SANS. USANS data before de-smearing are given in Figure S2.

The scattering from three separators were compared: Celgard polypropylene, polyvinylidene fluoride (PVDF) and quartz glass microfibre, with relatively strong scattering expected from all of them as a result of their porosity. Separators that have been wetted with electrolyte experience pore deformation[29] and modification of the SLD contrast at the pore surface; to capture data representative of separators in the *in situ* cell, data for each separator impregnated with approximately 200 µL electrolyte (1 M LiPF$_6$ in ethylene carbonate/dimethyl carbonate) in a sealed laminated pouch were collected and data for an empty pouch subtracted (Figures 3B and 3C) according to equation (9) of §7.3 of the Methods section. We note that scattering from the pouch is substantially lower than that from the electrolyte-wet separator (Figure S3A). Scattering from the electrolyte was measured but not subtracted from these data, noting negligible scattering from the electrolyte volume. Celgard exhibited substantial anisotropic scattering, likely as a result of directional specific pore deformation.[30] Although the differential scattering cross-section per unit volume was comparable between wet PVDF and glass microfibre (Figure 3C), scattering from wet glass microfibre per unit area was 1-2 orders of magnitude greater, depending on Q, than that from wet PVDF because of the 10 times larger thickness of the former. Consequently, beam transmission through wet glass microfibre was reduced by coherent scattering ($T_{SAS}$ = 37.1%) and therefore PVDF was selected as the separator for the *in situ* cell (Table S1).

SANS and USANS data were collected from a laminated aluminium pouch and lithium foil sealed within a laminated pouch. Data representative of lithium were obtained by subtracting the data for the pouch from that for lithium in the pouch (Figure 3D), resulting in statistically poorer data (Figure S3A). Data for lithium in the pouch contain two approximately linear regions of different slopes on a log-log scale, with $Q^{-3.69(4)}$ variation in the USANS region between $4\times10^{-5}$ and $4\times10^{-4}$ Å$^{-1}$ and $Q^{-3.07(5)}$ variation in the SANS region between $2\times10^{-3}$ and $4\times10^{-2}$ Å$^{-1}$. Data resemble that for other bulk metals,[26,27] where the $Q^{-3}$ variation at low $Q$ is typical of dislocation scattering and $Q^{-3.5}$ variation at high $Q$ corresponds to a mixture of dislocation scattering and scattering by pores, impurities or surfaces.

Two identical *in situ* pouch cells were produced and the reproducibility of the USANS signal before cycling confirmed (Figure S5D). The calculated coherent SLD value of each component is given in Table S2. Neutron transmission after attenuation from coherent scattering ($T_{SAS}$), absorption and incoherent scattering ($T_{A,I}$) and the three effects combined ($T_{A,I,C}$) were estimated from USANS data at a wavelength of 4.74 Å.[31,32] Transmission through select components and the *in situ* cell are given in Table S1. Multiple coherent scattering is negligible for the *in situ* cell before cycling ($T_{SAS}$ = 88.2%, where multiple scattering is considered significant for T$_{SAS}$ less than 80%[32]). Transmission through the *in situ* cell calculated from incoherent scattering and absorption ($T_{A,I}$) was approximately 72% before cycling, with the strongest attenuation by the laminated pouch, electrolyte-wet PVDF, and lithium foil, likely as a result of absorption by Li and large incoherent scattering by hydrogenated components.

Seidlmayer *et al.*[19] have suggested that battery components form distinct macroscopic layers such that the scattering from each is independent and that the data from a cell is composed of the sum from the individual components. The scattering from low surface area solids such as smooth copper and lithium foils, as well as the laminated pouch, is expected to differ substantially to that from pore surfaces of the PVDF, and we find decreased intensity of electrolyte wet PVDF relative to dry PVDF (Figure 3B and 3C) due to differences in SLD contrast at the interface ($\Delta\rho^2 = 8.41\times10^{-12}$ Å$^{-4}$ in air compared with $1.37\times10^{-12}$ Å$^{-4}$ in electrolyte). Consequently, we treat the electrolyte and PVDF as a single component in following discussions. Scattering per unit area from the *in situ* cell is compared in (Figure 4A) to that of the sum of the scattering per unit area from each component according to equation (8) of §7.3 of the Methods section developed in Technical Note S3. We find that the sum of component scattering reproduces that of the overall cell, confirming the independence of component contributions from each other and facilitating the subtraction of unwanted scattering in subsequent analyses.

The differential scattering cross-section per unit area obtained from equation (7) of §7.3 of the Methods section, representative of the relative scattering intensity from each component as opposed to data per unit volume, is shown for the individual components and the *in situ* cell in Figure 4A. Before cycling, lithium contributes significantly at $Q < 10^{-4}$ Å; however, in the remaining $Q$ range, the majority of scattering originates from the electrolyte-wet PVDF, by about an order of magnitude compared to that from lithium. Contribution from lithium foils to data in the USANS region is comparable to that of copper foils and that of the laminated pouch in the SANS range. A high incoherent background is observed at $Q > 10^{-2}$ Å$^{-1}$, arising from electrolyte-wet PVDF and the laminated pouch, as expected from their hydrogen-rich composition.

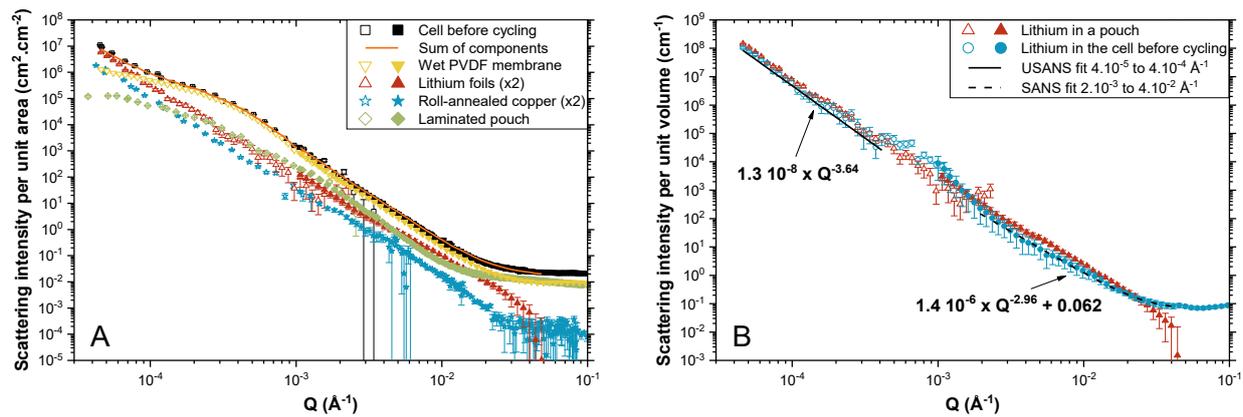

**Figure 4**: Desmeared USANS (open symbols) and SANS (closed symbols). A) Differential scattering cross-section per unit area from the *in situ* cell before cycling, individual components, and the sum of all components. B) Differential scattering cross-section per unit volume from lithium in the *in situ* cell before cycling after subtraction of scattering from inactive components and that from lithium in a pouch after subtraction of scattering from the pouch. The power-law exponent for lithium in the cell before cycling was extracted by fitting of the function $A \times Q^{-n}$ over smeared USANS data and $A \times Q^{-n} + B$ function over SANS data within the $Q$ ranges in legend. Refined values were $A = 1.3(12) \times 10^{-8}$ and $n = 3.64(10)$ for USANS and $A = 1.4(10) \times 10^{-6}$, $B = 0.062(16)$ and $n = 2.96(16)$ for SANS. Corresponding data before desmearing are shown in Figure S4.

The scattering from lithium in the *in situ* cell before cycling and after subtracting scattering from other components is shown in Figure 4B. The statistical precision of scattering for lithium before cycling is poor due to the relatively low scattering relative to that from inactive components Figure S3B, and future experiments may improve this by contrast-matching separator and electrolyte via deuteration. Scattering from lithium in the uncycled cell and for lithium in a pouch are comparable (Figure 4B), suggesting scattering does not originate from the surface, where a ten fold increase of scattering is expected from the change of SLD contrast at the interface upon contact with electrolyte ($\Delta\rho^2 = 0.67 \times 10^{-12}$ Å$^{-4}$ in argon compared with $6.5 \times 10^{-12}$ Å$^{-4}$ in electrolyte).

A small increase in intensity at approximately $Q = 10^{-3}$ Å$^{-1}$ is observed for lithium in the cell relative to lithium alone; however, statistical confidence of the feature is low. Data for lithium in the cell before cycling follows a $Q^{-3.64(10)}$ trend in the USANS region between $Q = 4 \times 10^{-5}$ and $4 \times 10^{-4}$ Å$^{-1}$, and a $Q^{-2.96(16)}$ trend in the SANS region between $Q = 2 \times 10^{-3}$ and $4 \times 10^{-2}$ Å$^{-1}$ (Figure 4B), comparable to those determined for lithium alone (Figure 3D). The similarity of scattering from lithium in the uncycled *in situ* cell and that for lithium alone suggest that lithium is not substantially affected by contact with electrolyte before cycling where, in the latter case, the lithium surface reacts with electrolyte to form a several nm thick SEI before cycling.[33,34]

## 3. Post cycling SANS and USANS

Two nominally identical symmetrical lithium metal *in situ* cells each underwent electrochemical cycling at different applied current density, which influences the rate at which lithium is deposited and extracted from electrode surfaces. Current densities of 2 mA/cm$^2$ and 20 mA/cm$^2$ were compared. At both current densities, the formation of dendritic lithium is expected to occur in reasonable time with the transition from mossy to dendritic lithium expected to occur earlier at the higher current density.[8] For each cell, a constant current was applied for 1 h followed by the collection of 3-5 h USANS measurements at open circuit voltage. One cell underwent two cycles (alternating "charge" and "discharge" processes) at 20 mA/cm$^2$ and the other cell underwent four consecutive "charges" followed by four consecutive "discharges" at 2 mA/cm$^2$, with a final "discharge" at 20 mA/cm$^2$. Applied current and measured voltage are shown in Figure 9. The calculated amount of lithium exchanged after each galvanostatic step is 5.2 ± 0.4 and 0.52 ± 0.04 mg/cm$^2$ at 20 and 2 mA/cm$^2$, respectively, and the calculated mass of lithium in each electrode is shown in Figure S11, assuming no loss. After cycling, both cells showed tighter compression against quartz slides, consistent with lithium volume expansion from increased surface porosity.[35]

Although the lithium surface on each side of the cell may differ following alternating lithium deposition and extraction, the data contain information from both electrodes. All scattering data are shown as a differential cross-section per unit volume and considers the constant initial volume of lithium in the cell (2 × 200 μm × beam area) ignoring volume changes. USANS data measured between each galvanostatic step for the cell cycled at 2 mA/cm$^2$ are shown in Figure 5A for the first four charges and Figure 5B for the following four 2 mA/cm$^2$ discharges and the discharge at 20 mA/cm$^2$. USANS data of the cell charged and discharged twice at 20 mA/cm$^2$ are shown in Figure 5C. Overall scattered intensity from the cell charged at 2 mA/cm$^2$ increases gradually during the first 3 charges with only minor variation post maximum, including after discharge at 20 mA/cm$^2$. At 20 mA/cm$^2$, the maximum scattering intensity is reached after the first cycle, with little variation thereafter. Both cells have a similar maximum scattering intensity, reaching an order of magnitude greater than before cycling, and the initial intensity is never recovered confirming the irreversibility of lithium surface transformations. Overall USANS data of the cell after galvanostatic cycling follow similar trends with an initial power-law decrease in intensity at $Q < 10^{-4}$ Å$^{-1}$ that becomes a gentler slope at intermediate $Q$ and a steeper slope at $Q > 10^{-3}$ Å$^{-1}$.

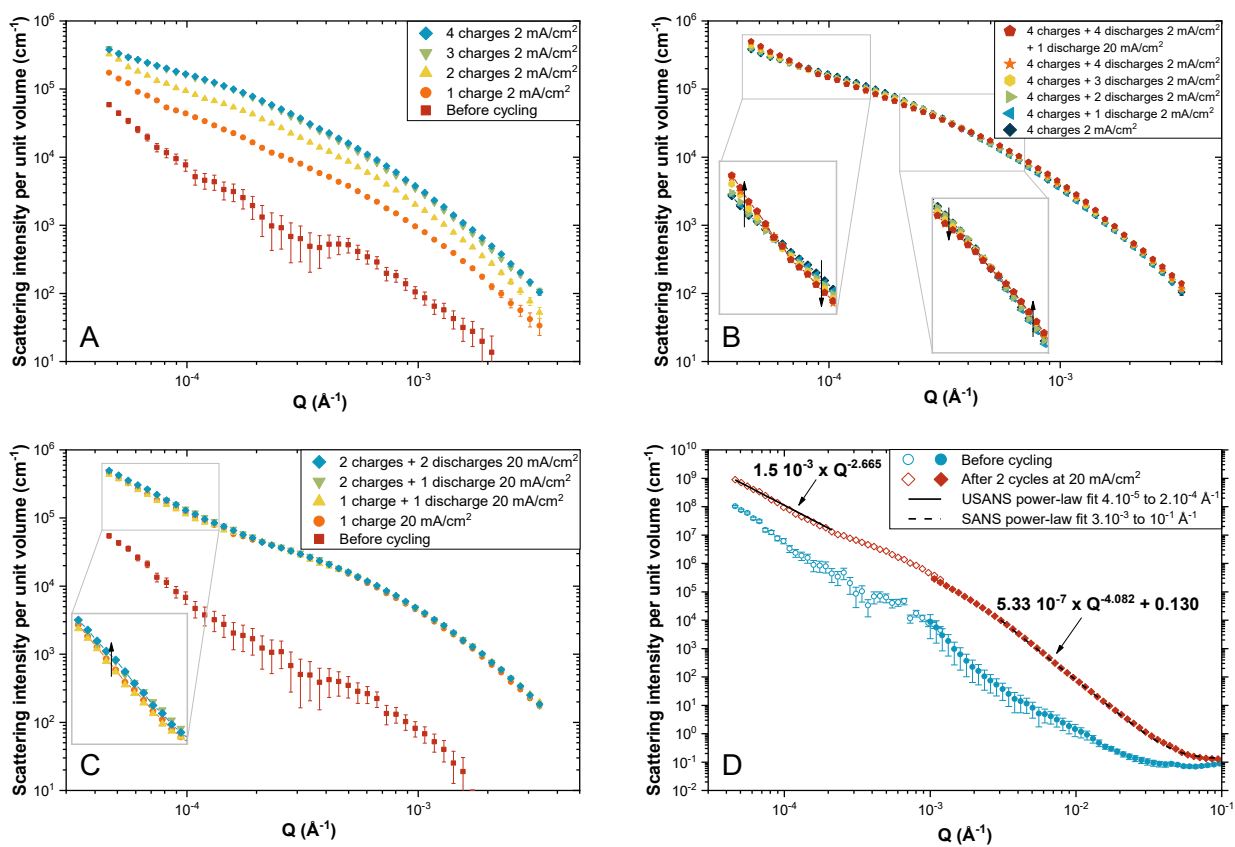

**Figure 5**: A) B) C) Slit-smeared USANS scattering shown as the differential cross-section per unit volume of lithium after subtraction of scattering from electrochemically inactive components from an *in situ* cell A) before cycling and after one, two, three and four consecutive "charges" at 2 mA/cm$^2$, and B) after four "charges" and followed by one, two, three and four consecutive "discharges" at 2 mA/cm$^2$, and an additional "discharge" at 20 mA/cm$^2$. C) From an *in situ* cell before cycling and after alternating "charges" and "discharges" at 20 mA/cm$^2$. Arrows are visual guides emphasizing intensity change. D) Desmeared USANS (open symbols) and SANS (closed symbols) scattering shown as a differential cross-section per unit volume for lithium in the cell before and after two cycles at 20 mA/cm$^2$. Power-law exponents for lithium in the cell after two cycles were extracted by fitting of the function $A \times Q^{-n}$ over smeared USANS data and $A \times Q^{-n} + B$ function over SANS data within $Q$ ranges given. Refined values were $A = 1.5(2) \times 10^{-3}$ and $n = 2.665(16)$ for USANS and $A = 5.33(18) \times 10^{-7}$, $B = 0.130(4)$ and $n = 4.082(7)$ for SANS. Corresponding raw data before desmearing are shown in Figure S5A, B, C, and Figure S6A.

USANS and SANS data for lithium in the cell before and after 2 cycles at 20 mA/cm$^2$ are shown in Figure 5D. Subtraction of inactive components does not substantially change the scattering pattern after cycling (Figure S3B) as a result of the order of magnitude increase of scattering from lithium after galvanostatic cycling; this larger increase than previously reported for *in situ* SANS data of lithium metal cells [20,21] is possibly a result of the higher relative current density. The USANS transmission through the cell decreased substantially after cycling Table S1 as a result of this increased scattering. Transmission after attenuation from coherent scattering ($T_{SAS} = 53\%$) suggests some degree of multiple scattering which may slightly decrease intensity at lower $Q$, noting a relatively small effect for samples with $T_{SAS} \approx 50\%$ measured on the same instrument.[32]

Data show a distinct change in shape before and after cycling (Figure 5D), with post cycling data exhibiting two approximately power-law decreases separated by a broad shoulder in the intermediate $Q$ range, having $Q^{-2.665(16)}$ slope in the USANS region from $Q = 4\times10^{-5}$ to $2\times10^{-4}$ Å$^{-1}$ and $Q^{-4.082(7)}$ slope in the SANS region from $Q = 3\times10^{-3}$ to $10^{-1}$ Å$^{-1}$. The $Q^{-4}$ variation at low $Q$ is consistent with smooth interfacial scattering and increased surface area post cycling, as observed with other techniques.[36,37] We postulate that scattering from lithium post cycling originates from the lithium–electrolyte interface; the scattering from the bulk foil volume, that dominates pre-cycling, is negligible in comparison.

## 4. Surface area from Porod's law

In non-particulate, non-uniform systems, scattering can originate from interfaces between volumes of different SLD, known as phases,[38,39] and we therefore consider that scattering from the cell post cycling originates from the lithium – electrolyte interface, with lithium metal and electrolyte taken as separate homogeneous volumes where SLD fluctuations are negligible. The $Q^{-4}$ slope at high $Q$ indicates a sharp change of SLD at the boundary,[26,38,39] and the increase of scattering in the cell post cycling is consistent with increased surface area.[36,37] The $Q^{-4}$ slope from the cell post cycling is sustained at $Q > 2\times10^{-3}$ Å$^{-1}$ ($Q^{-1} < 50$ nm), suggesting the absence of particles or porosity smaller than 50 nm, noting that features smaller than 100 nm are rarely observed for deposited lithium despite complex micrometre scale morphology shown by optical and scanning electron microscopy.[6,8-10,13,15]

In a non-particulate system with homogeneous regions of different SLD, the differential scattering cross-section per unit volume from "smooth" surfaces is expected to follow Porod's law at sufficiently high $Q$:[38,39]

$$\frac{d\Sigma}{d\Omega} = P \times Q^{-4} + B \tag{1}$$

with $B$ accounting for the background and the Porod exponent $P$ the contribution from all surfaces:

$$P = 2\pi \sum_i \left[(\Delta\rho_i)^2 \times \frac{S_i}{V}\right] \tag{2}$$

where $S_i$ is the surface area at interface i between two phases, $\Delta\rho_i$ is the SLD contrast between the phases each side of the interface, and $V$ is the volume occupied by the phases. The model can be extended to any number of phases.

The modelling of scattering from the lithium – electrolyte interface is complicated by the presence of the SEI – a component of complex and debated composition. The SLD of the SEI is probably intermediate between that of lithium ($\rho_L = -0.82 \times 10^{-6}$ Å$^{-2}$) and the electrolyte ($\rho_E = 1.73 \times 10^{-6}$ Å$^{-2}$), with neutron reflectometry suggesting a value $\rho_S \approx 0.8 \times 10^{-6}$ Å$^{-2}$,[40] or alternatively, the SLD of the SEI may be very close to that of either lithium or the electrolyte, as postulated in previous SANS experiments.[20] Where a uniformly thick SEI coats the lithium electrode, all surfaces are equivalent and consequently, the surface area in the three-phase system is 1.86 times that of the two-phase system, as developed in Technical Note S1. A Porod's law description of the differential scattering cross-section per unit volume of the cell after two cycles at 20 mA/cm$^2$ in the SANS range $2\times10^{-3} < Q < 10^{-1}$ Å$^{-1}$ (Figure S6B) yields P = 78.0(3)×10$^{-8}$ Å$^{-4}$·cm$^{-1}$ and B = 0.118(4) cm$^{-1}$, corresponding to a surface area per unit volume $S_V$ = 19.08(7)×10$^3$ cm$^2$/cm$^3$ in the two-phase system and $S_V$ = 35.57(14)×10$^3$ cm$^2$/cm$^3$ in the three-phase system.

Comparison with other experiments is not straightforward for cycled lithium. The differential scattering cross-section is usually divided by the irradiated volume to obtain the differential scattering cross-section per unit volume allowing comparison of the same material with variable thicknesses and beam area.[41] Following convention, scattering from lithium in the cell are presented per unit volume (in "absolute" units cm$^{-1}$), where data are divided by the lithium volume in the as prepared cell, containing 2×200 μm thick foils within the beam area. This approach assumes a uniform distribution of SLD heterogeneities within the sample volume; however, such a condition is not anticipated for lithium in the cell post cycling where scattering is conjectured to predominantly originate from the lithium – electrolyte interface, which is segregated to the lithium foil surface. In this condition, scattering intensity varies linearly with the volume of deposited lithium including pores, and not with the total volume including the excess, which depends on cell construction. This means that the differential scattering cross-section and, by extension, surface area, per unit volume cannot be compared between cells with different lithium foil thickness. This is further complicated by the unaccounted for volume expansion introduced by porous deposited lithium.[7,35]

A macroscopically uniform distribution of deposited lithium is expected across the electrode surface, where the approximate 6 mm distance between electrode edges and the area of the cell probed by the beam renders edge effects[42] likely negligible. Therefore, the magnitude of surface scattering is expected to vary linearly with sample area, similar to materials where scattering from a relatively homogeneous bulk is negligible in comparison to that from highly subdivided surfaces,[26,43,44] and where the differential scattering cross-section per geometric area (cm$^2$/cm$^2$) should be comparable between data for cells with the same surface condition but with different amount of excess foil. The conversion of surface area per unit volume $S_V$ to surface area per geometric area $S_A$ that considers both active surfaces is obtained using equation (12) in §7.3 of the Methods section by straightforward multiplication by 0.02 cm.

|  | Surface area per unit volume ($S_V$, $10^3$ cm²/cm³) | Surface area per unit mass ($S_M$, $10^3$ cm²/g) | Surface area per unit area ($S_A$, cm²/cm²) |
|---|---|---|---|
| **SANS Porod model** <br> 2 cycles 20 mA/cm² (2 phase) <br> 2 cycles 20 mA/cm² (3 phase) | 19.08(7) <br> 35.57(14) | 38.56(15) <br> 71.8(3) | 381.7(1.5) <br> 711(3) |
| **BET (Weber *et al.*)** <br> 1 cycle 1.2 and 0.48 mA/cm² <br> 4 cycles 1.2 and 0.48 mA/cm² <br> 10 cycles 1.2 and 0.48 mA/cm² | Not reported | 30 (foil) <br> 75 (foil) <br> 150 (foil) <br> 250 (powder) | 30 (foil) <br> 75 (foil) <br> 150 (foil) <br> 250 (powder) |
| **BET (Saito *et al.*)** <br> 1 h discharge 3 mA/cm² <br> 6 cycles 1 and 0.2 mA/cm² <br> 6 cycles 1 and 3 mA/cm² | Not reported | 25 <br> 132 <br> 258 | Not reported |
| **X-ray tomography (Taiwo *et al.*)** <br> 10 cycles <br> 70 cycles <br> 135 cycles | 0.05 <br> 0.4 <br> 0.6 | Not reported | 0.175 <br> 5.2 <br> 10.8 |

**Table 1**: Surface area of deposited lithium calculated from SANS data of the *in situ* cell after two cycles at 20 mA/cm² (Figure S6B) using Porod's law for two- and three-phase models, alongside the reported surface area of lithium metal in cells post cycling obtained using gas adsorption and X-ray tomography. Weber *et al.* report *ex situ* data for lithium after 10 cycles extracted from the cell as foil plated on copper or powder scraped from copper, indicated as "foil" and "powder", respectively. Standard uncertainties estimated from the least-square regression analysis are shown in parentheses.

The surface area determined from Porod's law is compared to quantitative measurements of surface area reported using *in situ* X-ray tomography[11,45] and *post mortem* gas adsorption using Brunauer-Emmett-Teller (BET) theory[36,37] in Table 1. Quantitative reports of the surface area of cycled lithium are scarce, with qualitative descriptions from microscopy measurements substantially more common. Taiwo *et al.* measured a surface area per unit volume of $0.05\times10^3$, $0.4\times10^3$ and $0.6\times10^3$ cm²/cm³ within 35, 130 and 180 µm-thick electrodeposited lithium foils using X-ray tomography, corresponding to surface area per unit area[11] of 0.175, 5.2 and 10.8 cm²/cm², respectively, and below that measured using SANS or BET gas sorption. *In situ* X-ray tomography severely underestimates lithium surface area as a consequence of its approximate 1 µm resolution limitation. Description of USANS data by the Debye-Anderson-Brumberger (DAB) model, described later, confirms that micrometric-scale lithium features contribute to surface area determined by tomography.

Only two reports of gas adsorption determined lithium surface areas are found,[36,37] perhaps due to the experimental need to extract foils from the cell, requiring washing and drying under an inert atmosphere, and unconventional use of argon as the adsorbent. A comparison of surface areas per unit mass $S_M$ excluding excess lithium is presented in Technical Note S2 and substantial differences are observed between SANS and BET and between BET reports. The surface area per unit area $S_A$ obtained from SANS here, less prone to thickness uncertainties, and by Weber *et al.* using gas adsorption are within an order of magnitude, with a smaller $S_A$ reported from gas adsorption, likely as a result of the limited probing of internal surface, seen in Figure 1B, as evidenced by the increase in surface area measured for pulverised samples.[37] Insufficient information was given by Saito *et al.* to enable the determination of surface area per unit area. We note experimental differences likely influencing the determined surface area of lithium between this work and that of Saito *et al.* and Weber *et al.* including differences in cycling protocol, electrolyte, cell construction, as well as the substantial amount of lithium remaining attached to the separator after extraction from the cell in the work of Weber *et al.* These results highlight the applicability of SANS for the direct and representative determination of lithium metal surface area during cycling however further experiments are needed to discriminate between the suitability of two- and three-phase models.

## 5. Debye-Anderson-Brumberger modelling

The slope at $Q < 10^{-4}$ Å$^{-1}$ and shoulder around $10^{-3}$ Å$^{-1}$ in the USANS data cannot be modelled by Porod's law and the Debye-Anderson-Brumberger (DAB) model was used to describe the data in the combined SANS and USANS region.[46,47] The DAB model considers a non-particulate multi-phase system characterised by a correlation length $L$ that is related to the average distance between interfaces, with differential scattering cross-section:[46]

$$\frac{d\Sigma}{d\Omega} = D \times \frac{L^3}{[1+(QL)^2]^2} \tag{3}$$

where $D$ is a scaling factor and $D/L$ is related to the surface area similarly to Porod's constant:[46]

$$D/L = 2\pi \sum_i \left[(\Delta\rho_i)^2 \times \frac{S_i}{V}\right] \tag{4}$$

where $S_i$ is the surface area at interface $i$ between two phases, $\Delta\rho_i$ is the SLD contrast between phases either side of the interface and $V$ is the volume occupied by all phases. Derivation of the surface area for two and three-phase systems follow those for Porod's law.

In this model, scattering at large $Q$ is $Q^{-4}$ consistent with Porod's law, and approaches a soft maximum at $Q = 1/L$, where the shoulder at approximately $10^{-3}$ Å$^{-1}$ yields a correlation length close to 100 nm. The slope at $Q < 10^{-4}$ Å$^{-1}$ in our data is attributed to scattering from SLD heterogeneities larger than $1/Q_{min} \approx 2$ μm, where $Q_{min}$ is the smallest experimentally accessible scattering wavevector. This slope can be modelled by a second DAB term with larger $L$.[46] To distinguish between the two contributions, the scaling factor and correlation length of lithium features of sizes $< 1$ μm and $> 1$ μm are denoted $D_{nano}$ and $L_{nano}$, and $D_{micro}$ and $L_{micro}$, respectively. The complete model used to describe the *in situ* data is therefore given by:

$$\frac{d\Sigma}{d\Omega} = D_{nano} \times \frac{L_{nano}^3}{[1+(QL_{nano})^2]^2} + D_{micro} \times \frac{L_{micro}^3}{[1+(QL_{micro})^2]^2} + B \tag{5}$$

with background parameter $B$. Assuming both micrometric and nanometric inhomogeneities arise from lithium-electrolyte interfaces, the surface area is given by the sum:

$$D_{nano}/L_{nano} + D_{micro}/L_{micro} = 2\pi \sum_i \left[(\Delta\rho_i)^2 \times \frac{S_i}{V}\right] \quad (6)$$

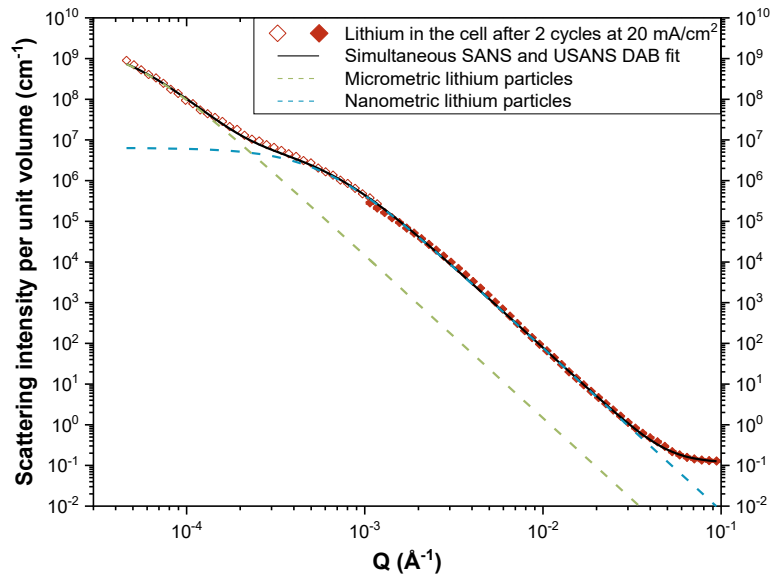

**Figure 6**: SANS (closed symbols) and desmeared USANS (open symbols) data of lithium in the *in situ* cell after 2 cycles at 20 mA/cm² and the corresponding DAB model calculation. The model was simultaneously refined against SANS and slit-smeared USANS data, with smearing applied to the model, as shown in Figure S6C, and refined parameters in Table S4.

The bimodal DAB model for two different sized inhomogeneities provides a reasonable description of USANS and SANS data for lithium in the cell post cycling at 20 mA/cm² (Figure 6). Refined model parameters (Table S4) reflect information from both electrodes, where lithium is alternatively deposited on one side and removed on the other, involving the partial redissolution of previously deposited lithium. The sum of $D_{nano}/L_{nano}$ and $D_{micro}/L_{micro}$ is 78.9(7)×10⁻⁸ Å⁻⁴·cm⁻¹, which compares well with the refined Porod parameter $P$ = 78.0(3)×10⁻⁸ Å⁻⁴·cm⁻¹, confirming the consistency of derived surface area using the two methods. 98% of the surface area arises from nanometric features as $D_{nano}/L_{nano} \gg D_{micro}/L_{micro}$, with the surface area arising from micrometric features close to $S_V$ = 0.4×10³ cm²/cm³ ($S_A$ = 8 cm²/cm²), comparable to the surface area obtained using *in situ* X-ray tomography (Table 1) of surface deposited lithium that is limited by resolution to micrometric porosity.[11,45] The refined correlation length corresponds to average distances between lithium/electrolyte interfaces separating statistically homogeneous volumes; one can envisage this to relate to the size of deposited lithium features and electrolyte-filled pores between them (Babinet's principle). $L_{nano}$ = 169.4(11) nm is similar to the width of so-called whiskers (Figure 1A) as well as the nanoporosity within mossy layers observed using electron microscopy.[6,13,15,48] $L_{micro}$ = 2.06(5) µm is similar to the size of microscopic pores within mossy lithium (Figure 1B) and to the width of dendrites seen using optical and electron microscopies.[6,8,9,13] $D_{nano}$ and $D_{micro}$ are quantitatively related to the lithium/electrolyte interfaces separating statistically homogeneous volumes at distances $L_{nano}$ and $L_{micro}$, respectively. As lithium volume expansion is not considered, $D_{nano}$ and $D_{micro}$ scale with the areal quantity of surfaces and not the volume concentration of surfaces conventionally expected.

The bimodal DAB model was fitted to USANS data post galvanostatic cycling, with B fixed to 0.117 cm$^{-1}$ in line with the negligible background, as shown in Figure S7 and refined parameters given in Table S5. Refined parameter values were in agreement between combined USANS + SANS data and USANS only data after 2 cycles at 20 mA/cm$^2$ (Table S4), with the exception of the $D_{nano}$ value, which was 10% larger when extracted using USANS data, as expected given the relative differences in $Q$ range. Generally, the USANS data were well described by the model, noting limited data from $Q = 4\times10^{-5}$ to $10^{-4}$ Å$^{-1}$ where $L_{micro}$ and $D_{micro}$ are determined, and where the signature for these features was not clearly present in data from the cell cycled at 2 mA/cm$^2$ after the 3$^{rd}$ and 4$^{th}$ charge, as well as the 1$^{st}$ and 2$^{nd}$ discharge. In these data, strong correlation between the $D_{micro}$ and $L_{micro}$ value (Figure S8A) was identified, and although convergence was achieved with good reproducibility of the refined $D_{micro}$ and $L_{micro}$ value, an unphysical substantial change in $L_{micro}$ was obtained between the 2$^{nd}$ and 3$^{rd}$ charges at 2 mA/cm$^2$. Correlations for all other refined parameters were negligible (Figure S8B). To account for correlations between $D_{micro}$ and $L_{micro}$, $L_{micro}$ was fixed to the average value of 2.1 µm as obtained from data that did not present these correlations, noting the inherent introduction of bias in the determined $D_{micro}$ as a result (Figure S8C and S8D), and a relatively smaller bias in the determined $D_{nano}$, $L_{nano}$, and $S_v$ value (Figure S9). Refinement results where $L_{micro}$ was fixed are presented in Figure S10 and Table S6.

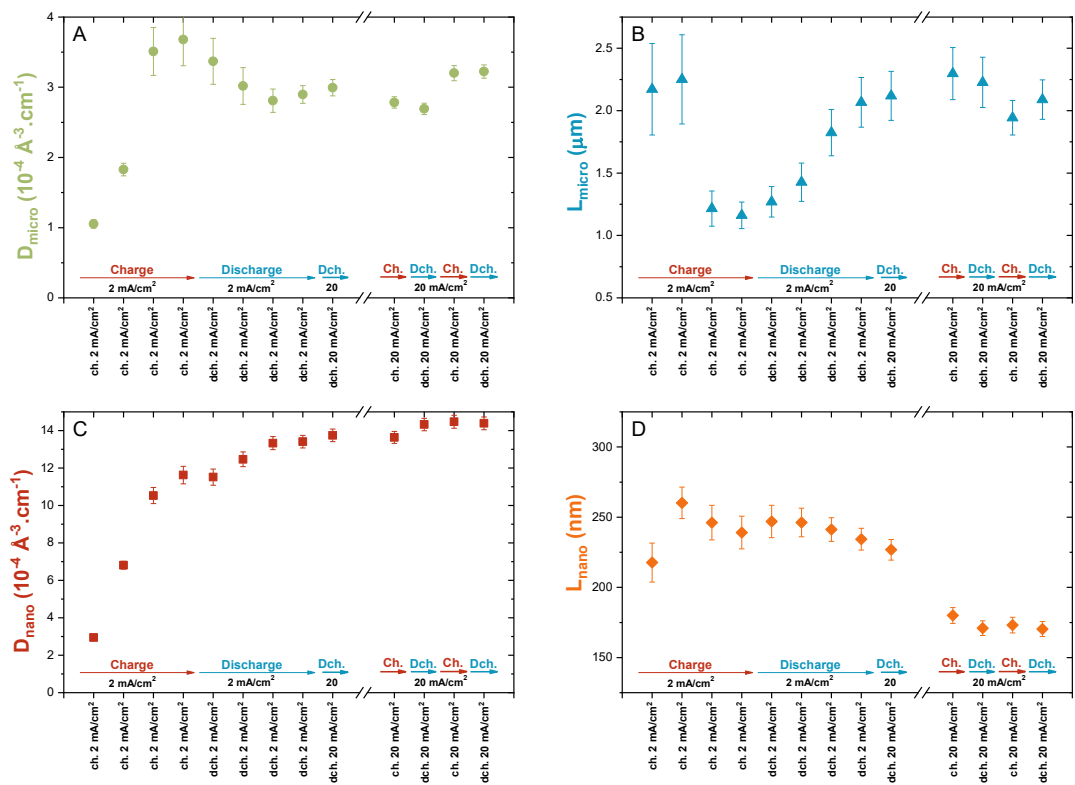

**Figure 7**: Refined parameters of the bimodal DAB model for lithium using USANS data of the *in situ* cell after each galvanostatic step, where "Ch." and "Dch." refer to "charge" and "discharge", respectively. A) Scale factor for micrometric features $D_{micro}$, B) correlation length $L_{micro}$, C) Scale factor for nanometric feature $D_{nano}$, and

D) correlation length $L_{nano}$. Error bars are standard uncertainties estimated from the least-square regression analysis.

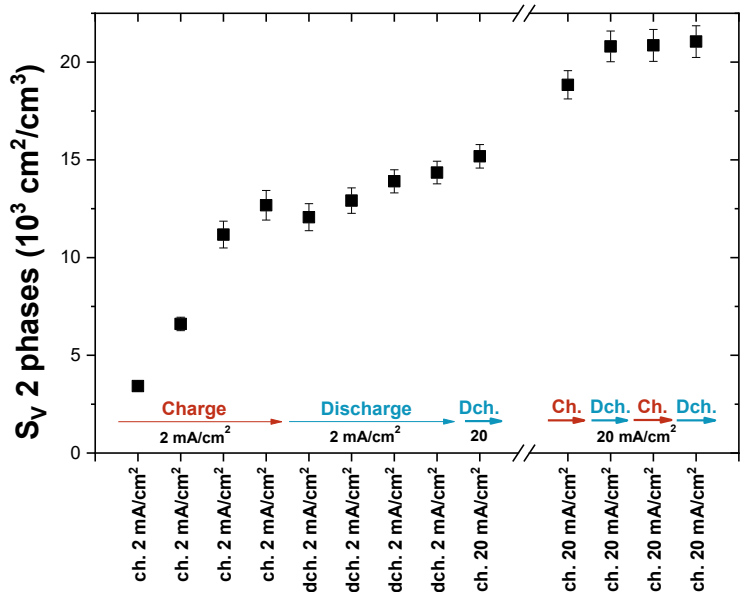

**Figure 8**: Refined surface area per unit volume $S_V$ considering two phases (lithium and the electrolyte) calculated from refined parameters of the bimodal DAB model using USANS data for lithium in the *in situ* cell after each galvanostatic step. Error bars are standard uncertainties estimated from the least-square regression analysis.

Refined parameters of the bimodal DAB model against USANS data with B fixed to 0.117 cm$^{-1}$ are plotted in Figure 7 and Figure 8. With increasing cycling, parameters for the cell cycled at 2 mA/cm$^2$ approached values obtained for the cell cycled at 20 mA/cm$^2$. $L_{nano}$ was approximately 240 and 170 nm for the cell cycled at 2 mA/cm$^2$ and 20 mA/cm$^2$, respectively, comparable to the width of lithium "whiskers" observed by electron microscopy[10,15,33], and where a reduction of whisker width was observed with increasing current density[15,48]. A relatively slow reduction of $L_{nano}$ was observed during the 4 h discharges at 2 mA/cm$^2$ and the following discharge at 20 mA/cm$^2$, whereas $L_{nano}$ remained relatively constant in the cell cycled at 20 mA/cm$^2$. The quantity of surface between nanometric features ($D_{nano}$) increased during the first 4 h charges at 2 mA/cm$^2$ and remained relatively constant during the subsequent discharge at 2 mA/cm$^2$ and at 20 mA/cm$^2$ past the initial charge capacity. In the cell cycled at 20 mA/cm$^2$, $D_{nano}$ remained relatively constant and comparable to the value eventually reached by the cell cycled at 2 mA/cm$^2$, suggesting the dissolution of deposited lithium is partially reversible. The refined size of micrometric features ($L_{micro}$) was between 1 and 2 µm, although further interpretation of the physical meaning of this is statistically limited. The quantity of surface around micrometric features ($D_{micro}$) increased after cycling at 2 mA/cm$^2$ noting limited statistical accuracy (Figure S9), to a value close to that obtained after 1 h at 20 mA/cm$^2$.

The surface area per unit volume $S_V$ depends exclusively on $D_{nano}$ and $L_{nano}$, with negligible influence from micrometric lithium. An increase in $S_V$ is observed from 3×10$^3$ to 12×10$^3$ cm$^2$/cm$^3$ after 1 to 4 h charge at 2 mA/cm$^2$, followed by a slower increase from 12×10$^3$ to 15×10$^3$ cm$^2$/cm$^3$ after 1 to 4 h discharge at 2 mA/cm$^2$ (Figure 8). A higher surface area of 20×10$^3$ cm$^2$/cm$^3$ is reached after just 1 h at 20 mA/cm$^2$ in the second cell, which remained constant in subsequent cycles. This corresponds to surface areas per unit area $S_A$ from 60 to 300 cm$^2$/cm$^2$ in the 2 mA/cm$^2$ cell and around 400 cm$^2$/cm$^2$ in the 20 mA/cm$^2$ cell. Cycling history had a major impact on the evolution of surface area, where 1 h at 20 mA/cm$^2$ caused a smaller increase in $S_V$ when the cell was previously cycled at 2 mA/cm$^2$ compared to 20 mA/cm$^2$, noting that capacity may have been limited by dendrite-induced short circuits.

## 6. Conclusions

We present a lithium metal battery (LMB) with relatively simple pouch construction suitable for *in situ* small angle neutron scattering (SANS) and ultra-small angle neutron scattering (USANS) studies of the structural development of deposited lithium within LMBs; characteristics can be evaluated with good precision and much less difficulty compared to other techniques such as X-ray tomography, microscopy or gas adsorption. We demonstrate the sensitivity of SANS and USANS to the development of lithium-electrolyte interfaces arising from lithium deposition, and quantify the surface area and average distance between these interfaces using relatively simple Porod's law and Debye-Anderson-Brumberger models applied to the SANS and USANS data, respectively. Complex variations of surface area and distance between interfaces were observed depending on the cell cycling history. This work paves the way for future investigations probing the influence of parameters such as current density, charge duration and alternating lithium deposition/dissolution processes on the surface area and interfacial distances within the deposited lithium; such information is necessary to address the limitations that lithium dendrite growth has on LMB technology application.

## 7. Methods

### 7.1. Battery components

Three separators were investigated: polypropylene (Celgard 2400 Polypore, 25 µm thickness, 40 nm pore size, 40% porosity), polyvinylidene difluoride (PVDF) (Immobilon-P, Merck, 110 µm thickness, 450 nm pore size, 70% porosity), and quartz glass microfibre (Whatman QM/A, Sigma-Aldrich, 450 µm thickness, 2.2 µm pore size). Four current collectors were investigated: nickel mesh (TOB New Energy, 180 µm aperture, 50 µm threads), copper mesh (99.9%, The Mesh Company, 204 µm aperture, 50 µm threads), electrodeposited copper foil with one rough and one smooth side (> 98%, MTI, 9 µm thickness), and roll-annealed copper foil smooth on both sides (99.9%, Goodfellow, 25 µm thickness). Current collectors and separators were dried overnight at 80 °C under vacuum before their introduction into an Ar glove box. The electrolyte was made by dissolving 1 M lithium hexafluorophosphate (LiPF$_6$) (99.99%, Sigma-Aldrich) in a 1:1 volume mixture of ethylene carbonate (EC) (anhydrous, 99%, Sigma-Aldrich) and dimethyl carbonate (DMC) (anhydrous, 99.7%, Sigma-Aldrich). Solvents were dried overnight under 4A molecular sieves prior to dissolution of LiPF$_6$ at room temperature for two days within an Ar filled glove box with < 1 ppm O$_2$ and H$_2$O. Lithium metal (99.9%, Goodfellow, 200 µm thickness) was used as electrodes. Although lithium was stored in an Ar filled glove box with < 1 ppm O$_2$ and H$_2$O, a thin and uneven coating of a black or white crust as a result of oxidation or nitridation, respectively, was present on the lithium metal, which was removed by abrasion using a rough polypropylene block until a smooth metallic surface was obtained. Aluminium laminated film (MTI, 115 μm thickness), referred to as laminated pouch, was used for isolating air-sensitive battery components, consisting of an inner polypropylene layer facing battery components and an outer nylon 6,6 polyamide layer, with the two layers encasing a layer of aluminium metal, attached with adhesive of unknown composition (not provided by manufacturer).

SANS and USANS data from individual battery components were collected to quantify their contribution to the overall scattering signal from the cell and to inform the construction of a symmetrical pouch cell favourable to the observation of changes in the deposited lithium. Laminated Al pouch, separators, and current collectors were handled in air. Lithium metal and separators wetted with approximately 200 μL electrolyte were sealed in the laminated Al casing within an Ar glove box. Components, aside from the electrolyte, were maintained flat between two quartz slides and taped to a flat sample holder during SANS and USANS measurements (Figure 2D). Electrolyte was introduced between two quartz slides separated by 300 μm and sealed with a compressed O-ring during USANS measurements and introduced into a quartz cuvette (Hellma cell) of 1 mm thickness and sealed by a Teflon cap for SANS measurements. Quartz slides and cuvette scattering measurements were also made and formed part of the empty cell measurements used in the background subtraction during data processing.

7.2 Preparation of symmetrical cells and electrochemical cycling for *in situ* SANS and USANS

Two symmetrical lithium metal pouch cells were prepared in an Ar filled glove box with < 1 ppm $O_2$ and $H_2O$. Lithium foil electrodes 2.5 cm × 2.5 cm, comprising approximately 10 mg/cm$^2$ (38.6 mAh/cm$^2$), were placed on 3.0 cm × 4.0 cm roll-annealed copper current collectors and aligned on each side of a PVDF separator wetted with approximately 200 μL of 1 M $LiPF_6$ in EC/DMC electrolyte. Measurements of cut lithium square electrodes post abrasion showed a thickness of 200 ± 10 μm, edge length of 2.5 ± 0.1 cm and initial mass 62 ± 4 mg. The assembly was sealed in a laminated aluminium pouch and electrical connections made with Ni tabs mechanically welded to current collectors. A representation of the cell is shown in Figure 2A, 2B and 2C.

Batteries were maintained flat between quartz slides with a slight pressure applied by bulldog clips to prevent misalignment of electrodes. Each battery underwent a different electrochemical cycling program, summarised in Figure 9. Although the terms "charge" and "discharge" have no real meaning in a symmetrical cell where lithium is in excess, they are used here to indicate when the direction of applied current is reversed. One battery underwent two "charge" and "discharge" cycles at a current density of 20 mA/cm$^2$ while the other underwent four consecutive "charge" followed by four consecutive "discharge" processes at 2 mA/cm$^2$, followed by one further "discharge" at 20 mA/cm$^2$. Each galvanostatic step was applied for 1 h using a PG302N (Autolab) potentiostat/galvanostat and the circuit was left open for 3-5 h during which USANS data were measured. SANS data of the 20 mA/cm$^2$ battery were measured prior to and following cycling in USANS studies.

The mass of lithium $m_{Li}$ exchanged between electrodes after a current $I$ is applied for a time $t$ is calculated by $m_{Li} = \frac{I \times t}{F} \times M_{Li}$ where F = 96485 A.s/mol is the Faraday constant and $M_{Li}$ = 6.94 g/mol is the molar mass of lithium. The mass of lithium at each "side" of the battery as bulk lithium foil, and as deposited from electrochemical processes, is plotted in Figure S11 assuming this as a total inventory with no loss from side reactions and the complete reinsertion of previously deposited lithium on current reversal.

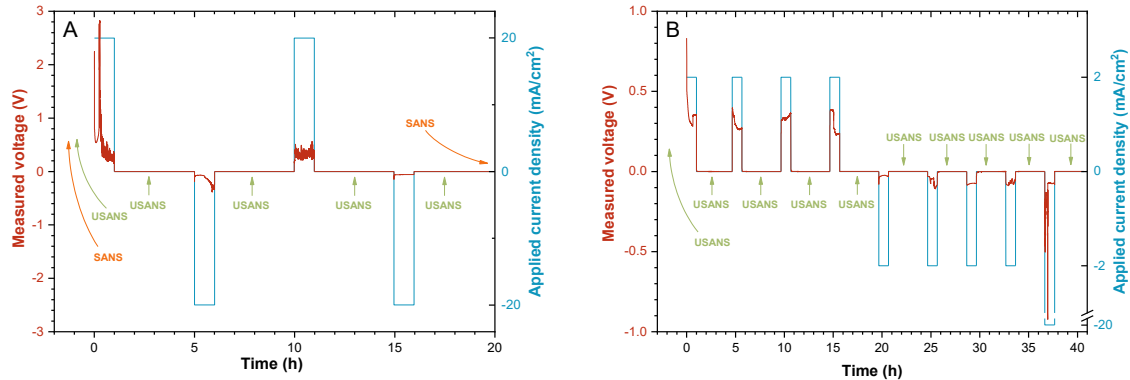

Figure 9: Applied current density to, and measured voltage of, symmetrical lithium cells at A) 20 mA/cm² and B) 2 mA/cm² with a final 20 mA/cm² galvanostatic step. Green arrows indicate the approximate time for USANS measurements between steps and orange arrows mark SANS measurements of the cell before and after cycling at 20 mA/cm².

7.3. Small and ultra-small angle neutron scattering measurements

USANS data (approximate $Q$ range = 3.5x10$^{-5}$–10$^{-2}$ Å$^{-1}$) were measured on the slit-geometry instrument Kookaburra at the Australian Nuclear Science and Technology Organisation (ANSTO)[22] in high flux mode with a neutron wavelength of 4.74 Å and vertical resolution parameter of 0.0586 Å$^{-1}$. The beam aperture diameter was 29 mm for individual components and 12 mm for cells. Multiple scattering was estimated from USANS data using the beam transmission $T_{SAS}$ method[31,32] using neutron counts measured with and without the sample to estimate neutron transmission through the sample after attenuation by coherent, incoherent scattering, and absorption combined ($T_{A,I,C}$), by absorption and incoherent scattering combined ($T_{A,I}$), and by coherent scattering only ($T_{SAS}$), the latter being an indication of multiple scattering.

SANS (6 x 10$^{-4}$ < $Q$ / Å$^{-1}$ < 0.6) data were measured on the pinhole-geometry instrument Quokka at ANSTO[23] in four configurations: at a neutron wavelength of 5.0 Å and source-to-sample distances of 20.1, 8.0, and 1.3 m, and at a wavelength of 8.1 Å and source-to-sample distance of 20.1 m using MgF$_2$ focussing optics). Quokka features a 1 x 1 m² area detector which was used to identify anisotropic scattering from individual cell components.

Experimentally measured scattering intensity was converted to differential macroscopic scattering cross-section per unit volume (absolute calibration) using empty-beam and direct beam attenuation measurements, and taking into account the thickness of components given in §7.1 of the Methods section.[41,49] For each component i, the differential scattering cross-section per unit area $I_A$ (in cm².cm$^{-2}$) and per unit volume $I_V$ (in cm$^{-1}$) are related by the thickness t (in cm) of the component:

$$I_A(i) = I_V(i) \times t(i) \qquad (7)$$

When a sample contains several components i (sample = $\sum i$) and component scattering is independent from each other as demonstrated for those comprising the *in situ* cell except for separator and electrolyte treated together, scattering intensities per unit area are additive:

$$I_A(\sum i) = \sum_i I_A(i) \qquad (8)$$

And scattering per unit volume is obtained after multiplication by the thickness:

$$I_V(\sum i) \times \sum_i t(i) = \sum_i [I_V(i) \times t(i)] \qquad (9)$$

The calculation of scattering for lithium in the *in situ* cell is provided as an example in Technical Note S3.

Data were subsequently processed with python scripts in Mantid[50] for additions, subtractions, and, in the case of USANS data, desmearing. Data additions and subtractions were performed independently for SANS and slit-smeared USANS after a spline interpolation of the data to a common $Q$ grid of evenly spaced points on a log-scale. For plotting purposes, USANS data were desmeared as the last step using the Lake algorithm.[51] Differential scattering cross-sections are given with solid angles expressed in units of steradians but steradians were omitted from unit labels by convention.[52] In plot axis labels, "differential scattering cross-section" is simplified to "scattering intensity". Model fitting was done using the program SASView 5.0.5 for slit-smeared USANS and/or SANS data with horizontal slit-smearing applied to the model when USANS data were included. Standard uncertainties of refined parameters are estimated from the least-square regression analysis and noted following recommendations of the International Union of Crystallography.[53]

For calculations of surface areas, the coherent neutron scattering length density (SLD) was calculated using $SLD = \frac{N_A \rho}{M} \times \sum_i p_i b_{C,i}$ where $N_A$ is Avogadro's number, $\rho$ is the bulk density, $M$ is the molar mass, $b_{C,i}$ are the atomic coherent scattering lengths[54] of the $i$ element in atomic proportion $p_i$. The surface area per unit volume $S_V$ (cm$^2$/cm$^3$) extracted from the coherent scattering cross-section per lithium volume corresponds to the surface contributing to scattering $S_i$ divided by the initial volume of lithium introduced in the cell:

$$S_V = \frac{S_i}{initial\ Li\ volume} = \frac{S_i}{beam\ area \times foil\ thickness \times number\ of\ foils} \qquad (10)$$

with two 200 µm-thick lithium foils in our experiment, neglecting thickness changes after cycling. The surface area per unit mass $S_M$ (cm$^2$/g) is derived from $S_V$:

$$S_M = \frac{S_i}{initial\ Li\ mass} = \frac{S_i}{beam\ area \times mass\ loading \times number\ of\ foils} = \frac{S_V \times foil\ thickness}{mass\ loading} \qquad (11)$$

with a mass loading of approximately 9.9 ± 1.4 mg/cm$^2$ in our experiment. Both $S_V$ and $S_M$ average the surface of the lithium – electrolyte interface over the total lithium inventory, including excess lithium, precluding comparison between cells with different amounts of lithium. Since surface scattering scales with sample area rather than volume,[26] the surface area per unit area $S_A$ (cm$^2$/cm$^2$), where only active lithium surfaces are considered, can be derived from $S_V$, considering one active electrode surface on each foil:

$$S_A = \frac{S_i}{active\ Li\ surface\ in\ the\ beam} = \frac{S_i}{beam\ area \times number\ of\ foils} = S_V \times foil\ thickness \quad (12)$$

## Acknowledgments

Access to Kookaburra and Quokka instruments was supported by ANSTO beamtime awards (proposal P8690 and DB9219). This work benefited from the use of the SasView program, originally developed under NSF award DMR-0520547. SasView contains code developed with funding from the European Union's Horizon 2020 research and innovation programme under the SINE2020 project, grant agreement No 654000. Authors are grateful to Liliana de Campo for troubleshooting issues with using the SasView program and useful discussions.

# Supplementary information file

# Direct *in situ* determination of the surface area and structure of deposited metallic lithium within lithium metal batteries using ultra small and small angle neutron scattering

Christophe Didier[1], Elliot P. Gilbert[1], Jitendra Mata[1] and Vanessa Peterson[1]*

[1] *Australian Centre for Neutron Scattering, Australian Nuclear Science and Technology Organization, Locked Bag 2001, Kirrawee DC, NSW 2232, Australia*

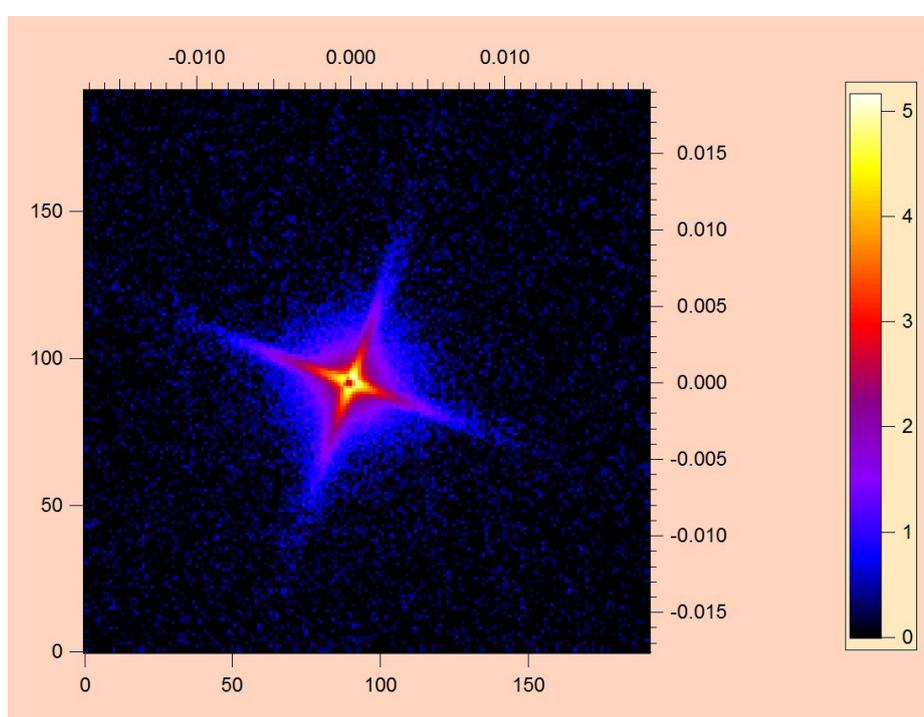

**Figure S1**: Anisotropic two-dimensional scattering data for nickel mesh. Intensity given in arbitrary units on logarithmic scale; positional information both in pixel number on 192 × 192 pixelated detector as well as $Q$ value. A similar pattern was obtained for copper mesh.

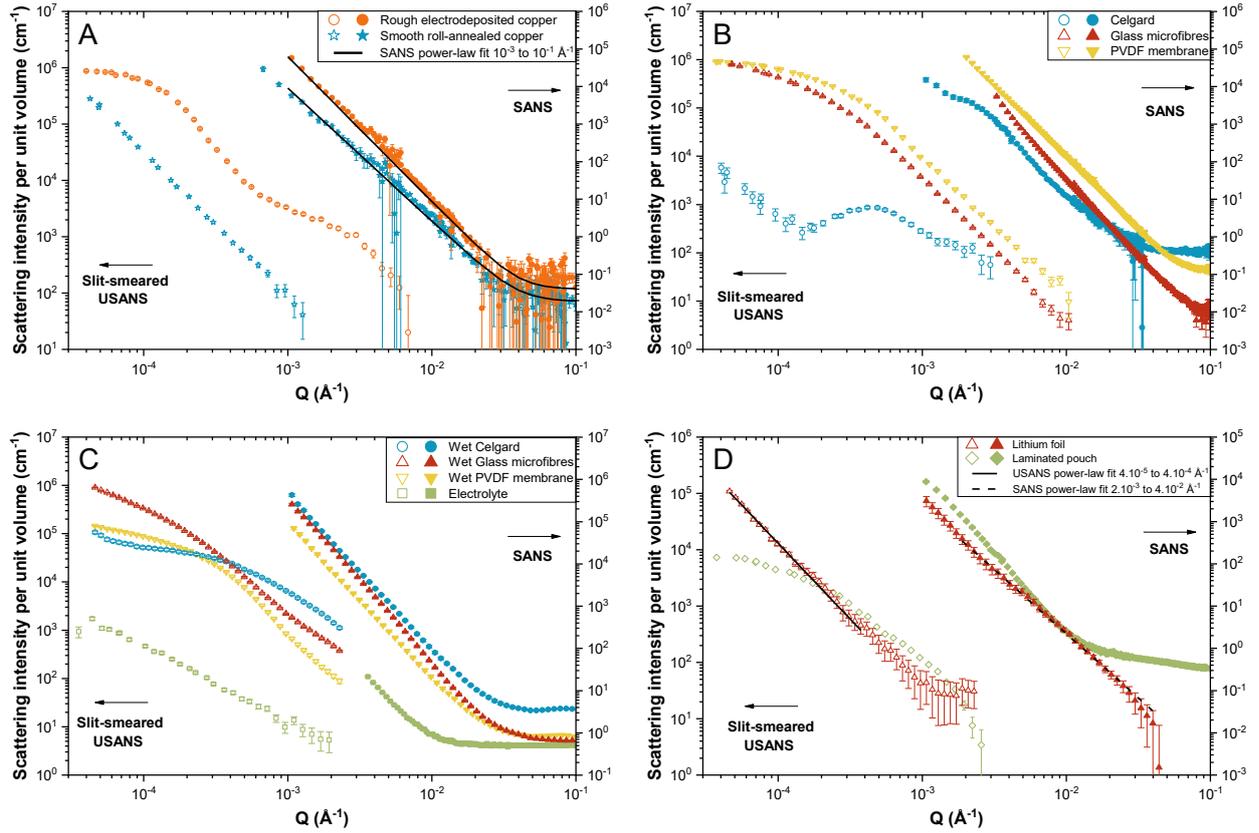

**Figure S2**: Differential scattering cross-section per unit volume for slit-smeared USANS (open symbols) and SANS (closed symbols) of A) rough electrodeposited and smooth roll-annealed copper foils with fitted power-law function $A \times Q^{-n} + B$ for smooth and rough copper over the SANS data between $10^{-3} < Q < 10^{-1}$ Å$^{-1}$. Refined values were $A = 1.66(14) \times 10^{-7}$, $B = 0.04(6)$ and $n = 3.858(15)$ for rough foil and $A = 2.06(19) \times 10^{-7}$, $B = 0.0188(14)$ and $n = 3.548(18)$ for smooth foil. B) Celgard, glass microfibres and PVDF membrane measured in air, C) electrolyte-wet Celgard, glass microfibres, PVDF membrane, and the 1 M LiPF$_6$ in EC/DMC electrolyte, shown after subtraction of scattering from the laminated pouch, and D) lithium metal after subtraction of scattering from the laminated pouch and laminated pouch data, and fitted power-law function $A \times Q^{-n}$ for lithium metal over SANS and smeared USANS data within $Q$ ranges shown in legend. Refined values were $A = 5(2) \times 10^{-9}$ and $n = 3.69(4)$ for USANS and $A = 1.6(4) \times 10^{-6}$ and $n = 3.07(5)$ for SANS.

| Component | $T_{A,I,C}$ (%) | $T_{A,I}$ (%) | $T_{SAS}$ (%) |
|---|---|---|---|
| Laminated aluminium pouch | 90.2 | 91.8 | 98.2 |
| Two roll-annealed copper foils | 96.2 | 99.2 | 97.0 |
| Two lithium foils | 70.8 | 83.7 | 84.6 |
| PVDF membrane with electrolyte | 87.8 | 90.4 | 97.1 |
| *In situ* cell before cycling | 63.2 | 71.7 | 88.2 |
| *In situ* cell after 2 cycles 20 mA/cm$^2$ | 37.1 | 70.0 | 53.0 |

**Table S1**: Neutron transmission through individual cell components and the *in situ* cell before cycling after attenuation from absorption, incoherent and coherent scattering ($T_{A,I,C}$), from absorption and incoherent scattering ($T_{A,I}$), and from coherent scattering only ($T_{SAS}$), the latter providing a measure of multiple scattering.

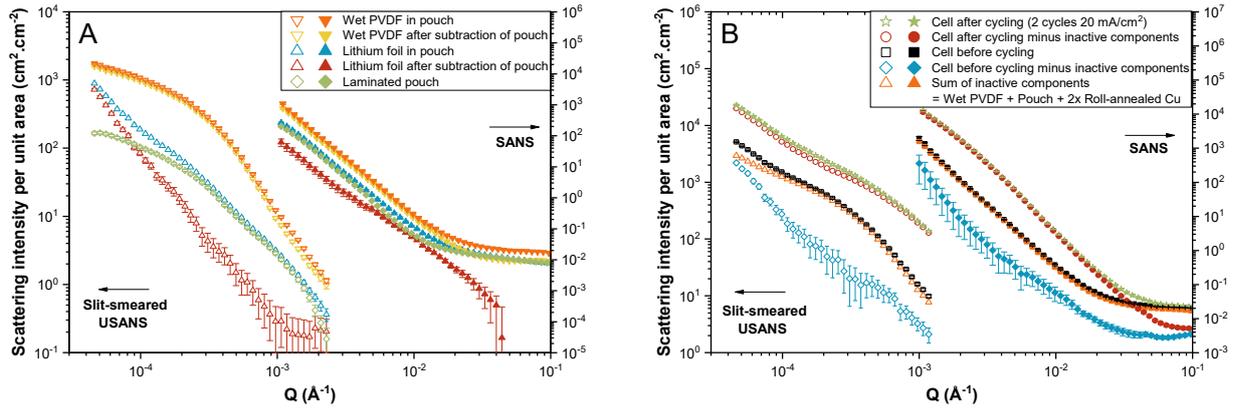

**Figure S3**: Scattering cross-section per unit area for slit-smeared USANS (open symbols) and SANS (closed symbols) showing the impact of component subtraction on data statistics. Data were first cropped and interpolated to a common x-axis to allow subtractions. Data shown for the A) electrolyte-wet PVDF membrane sealed in pouch and lithium foil sealed in pouch before and after subtraction of pouch data, B) *in situ* cell before cycling and after two cycles at 20 mA/cm², before and after subtraction of data from all inactive components which include electrolyte-wet PVDF, two smooth copper foils and the laminated pouch.

| Component | Coherent SLD ($10^{-6}$ Å$^{-2}$) | SLD contrast with respect to next component $\Delta\rho^2$ ($10^{-12}$ Å$^{-4}$) |
|---|---|---|
| Laminated pouch inner layer | -0.34 | 47.89 |
| Roll-annealed copper foil | 6.58 | 54.76 |
| Lithium foil | -0.82 | 6.5 |
| Electrolyte | 1.73 | 1.37 |
| PVDF | 2.90 | (repeats from centre of symmetrical cell) |

**Table S2**: Calculated coherent scattering length density (SLD) of the individual components in the *in situ* cell and the SLD contrast *Δρ²* between component surfaces, where contact between solids in the cell may not be continuous, with argon gas or electrolyte filling empty spaces.

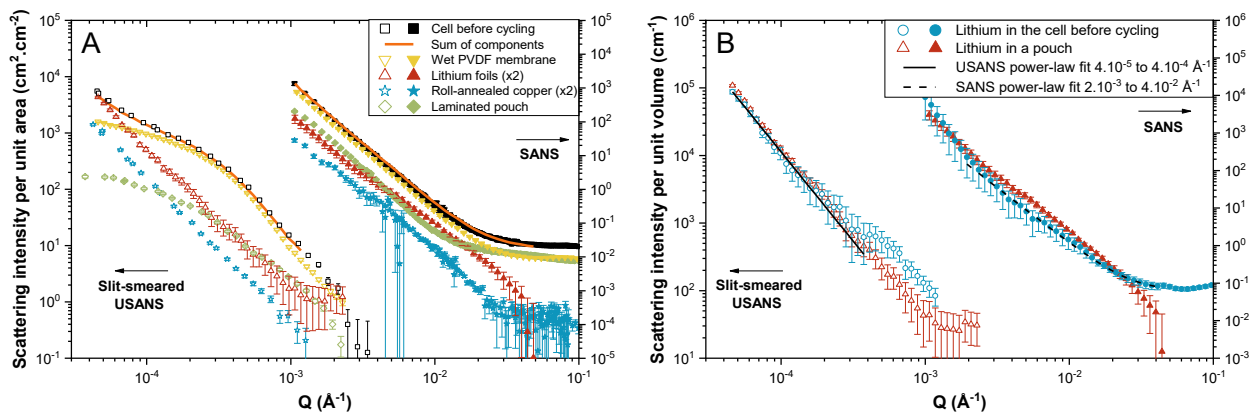

**Figure S4**: Slit-smeared USANS (open symbols) and SANS (closed symbols). A) Scattering cross-section per unit area from the *in situ* cell before cycling, each individual component and the sum of scattering for all components. B) Scattering cross-section per unit volume for lithium in the *in situ* cell before cycling after subtracting scattering from inactive components and that from lithium in a pouch after subtracting scattering from the pouch, and fitted power-law functions $A \times Q^{-n} + B$ over SANS and $A \times Q^{-n}$ over smeared USANS data within $Q$ ranges shown in legend for lithium metal in the cell before cycling. Refined values were $A = 1.3(12) \times 10^{-8}$ and $n = 3.64(10)$ for USANS and $A = 1.4(10) \times 10^{-6}$, $B = 0.062(16)$ and $n = 2.96(16)$ for SANS.

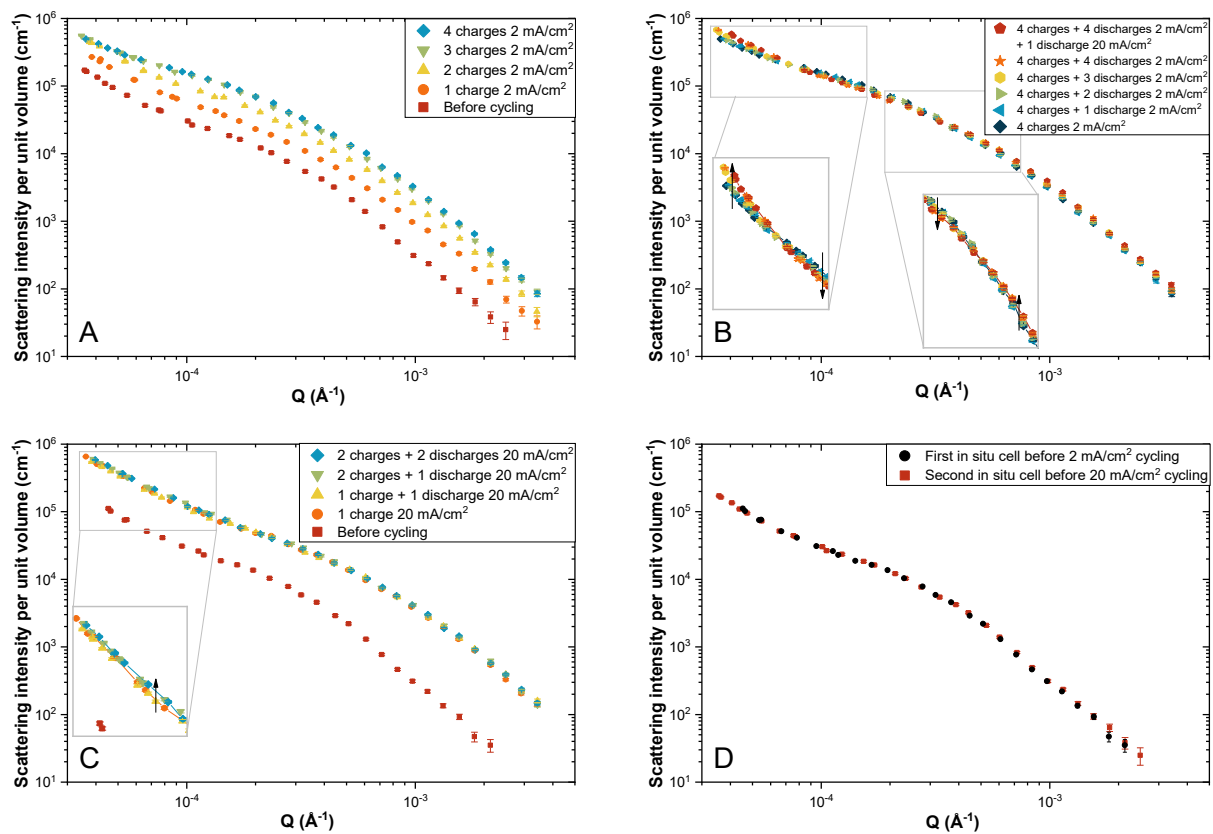

**Figure S5**: *In situ* slit-smeared USANS data as measured, before subtraction of inactive components for A) the first cell before cycling and after the first four 1 h charges at 2 mA/cm², B) the same cell subsequently discharged four times at 2 mA/cm² for 1 h and further discharged at 20 mA/cm² for 1 h, C) the second cell before cycling and after repeated cycles at 20 mA/cm². D) Overlay of data for the first and second cells before cycling, showing reproducibility.

**Technical Note S1**: Development of Porod exponent equations to determine the ratio between surface areas under the two- and three- phase assumptions

The three-phase system consists of two interfaces, one between lithium and the SEI, of surface $S_{L/S}$ and contrast $(\rho_L - \rho_S)^2$, and one between the SEI and the electrolyte, of surface $S_{S/E}$ and contrast $(\rho_S - \rho_E)^2$. The two-phase system consists of the interface between lithium metal and electrolyte of surface $S_{L/E}$ and contrast $(\rho_L - \rho_E)^2$. Expressions for the Porod exponent become:

$P = 2\pi/V \times (\rho_L - \rho_E)^2 S_{L/E}$ in the two-phase system
$P = 2\pi/V \times [(\rho_L - \rho_S)^2 S_{L/S} + (\rho_S - \rho_E)^2 S_{S/E}]$ in the three-phase system

Where a uniformly thick SEI coats the lithium electrode, all surfaces are equivalent with $S_{L/E} = S_{L/S} = S_{S/E} = S$. The surface area per unit volume S/V is denoted $S_V(2)$ and $S_V(3)$ in the two- and three-phase system, respectively, and equations are rearranged:

$S_V(2) = \dfrac{P}{2\pi \times (\rho_L - \rho_E)^2}$ in the two-phase system

$S_V(3) = \dfrac{P}{2\pi \times [(\rho_L - \rho_S)^2 + (\rho_S - \rho_E)^2]}$ in the three-phase system

Phase SLD are introduced into equations, with $\rho_L = -0.82 \times 10^{-6}$ Å$^{-2}$, $\rho_E = 1.73 \times 10^{-6}$ Å$^{-2}$ and $\rho_S \approx 0.8 \times 10^{-6}$ Å$^{-2}$ and after simplification:
$S_V(2) = P/[2\pi \times 6.5025 \times 10^{-12}]$ in the two-phase system
$S_V(3) = P/[2\pi \times 3.4893 \times 10^{-12}]$ in the three-phase system

Dividing the two, a ratio between surface areas under two- and three- phase assumptions is obtained:

$\dfrac{S_V(3)}{S_V(2)} = \dfrac{6.5025}{3.4893} \approx 1.86$

**Technical Note S2**: Comparison of surface areas per unit mass obtained from SANS Porod's law and reported from literature using BET

Weber *et al.* examined lithium deposited on copper and the $LiNi_{0.5}Mn_{0.3}Co_{0.2}O_2$ electrode in 1 M $LiPF_6$ in fluoroethylene carbonate : diethylcarbonate electrolyte. They report an increase in lithium surface area following 10 cycles at 0.48 (charge) and 1.2 (discharge) mA/cm$^2$ from $30 \times 10^3$ to $150 \times 10^3$ cm$^2$/g, and a surface area per unit area ranging between 30 and 150 cm$^2$/cm$^2$ with approximately 0.001 g/cm$^2$ lithium deposited at the charged state. The surface area remained stable after 10 cycles, with significant loss of capacity. Saito *et al.* cycled cylindrical cells with 150 μm thick lithium and $V_2O_5/P_2O_5$ electrodes in 1.5 M $LiAsF_6$ in ethylene carbonate : 2-methyltetrahydrofuran electrolyte. They report a surface area per unit mass of $25 \times 10^3$, $132 \times 10^3$, and $258 \times 10^3$ cm$^2$/g after 1 h discharge at 3 mA/cm$^2$, 6 cycles at 3 (discharge) and 1 mA/cm$^2$ (charge), and 6 cycles at 0.2 (discharge) and 1 mA/cm$^2$ (charge), respectively. In this work lithium was treated in dry air with < 10000 ppm humidity, noting that lithium reacts with nitrogen likely changing the surface.

The surface area per unit mass $S_M$ from SANS was calculated for comparison with BET data using equation (11) from §7.3 of the Methods section and reported in Table 1 considering the total inventory of active lithium. The mass of excess lithium was obtained from the cell capacity assuming no loss, with an approximately 75% excess lithium in our *in situ* cell (Figure S11), 90% in the cell of Saito *et al.*, and 0% in the "anode-free" cell of Weber *et al.*; this approach inevitably introduces uncertainty due to the unaccounted for irreversible capacity loss arising from loss of lithium inventory, electrolyte reduction, and short-circuiting induced by dendrites piercing the separator. Reports of the surface area per unit mass considering only deposited lithium (Table S3) therefore differ between Saito *et al.*, Weber *et al.* and this work.

|  | Surface area per unit mass: redeposited lithium ($S_M$, $\times 10^3$ cm$^2$/g) |
|---|---|
| **SANS Porod model** | |
| **2 cycles 20 mA/cm$^2$ (2 phases)** | 154.2(6) |
| **2 cycles 20 mA/cm$^2$ (3 phases)** | 287.4(11) |
| **BET (Weber *et al.*)** | |
| **1 cycle 1.2 and 0.48 mA/cm$^2$** | 30 |
| **4 cycles 1.2 and 0.48 mA/cm$^2$** | 75 |
| **10 cycles 1.2 and 0.48 mA/cm$^2$** | 150 |
| **BET (Saito *et al.*)** | |
| **1 h discharge 3 mA/cm$^2$** | 250 |
| **6 cycles 1 and 0.2 mA/cm$^2$** | 1 320 |
| **6 cycles 1 and 3 mA/cm$^2$** | 2 580 |

**Table S3**: Gravimetric surface areas normalised to the mass of redeposited lithium only, considering 75%, 0%, and 90% lithium excess in our study and that of Weber *et al.* and Saito *et al*. Standard uncertainties estimated from the least-square regression analysis are shown in parentheses.

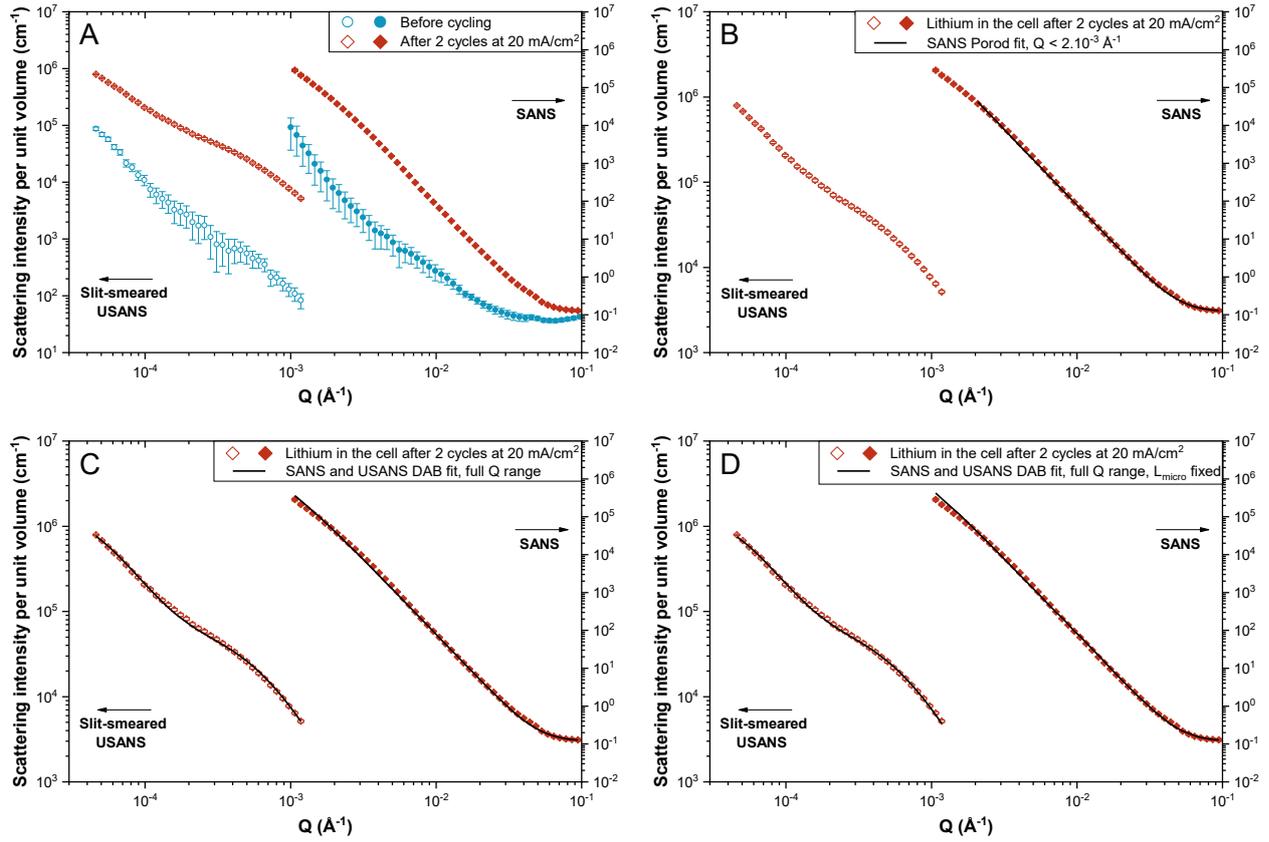

**Figure S6**: Slit-smeared USANS (open symbols) and SANS (closed symbols) scattering cross-section per unit volume A) for lithium in the cell before and after two cycles at 20 mA/cm$^2$. B) Lines show the refined Porod model alongside SANS data at $Q < 10^{-2}$ Å$^{-1}$. C) Lines show the refined bimodal DAB model alongside slit-smeared USANS data, with smearing applied to the model and SANS data. D) Lines show the refined bimodal DAB model alongside slit-smeared USANS data, with smearing applied to the model and SANS data, where L$_{micro}$ is fixed to 2.1 μm.

| Dataset | SANS + USANS | USANS | SANS + USANS | USANS |
|---|---|---|---|---|
| **Constraints** | All refined | B fixed | L$_{micro}$ fixed | B, L$_{micro}$ fixed |
| **B (cm$^{-1}$)** | 0.117(4) | 0.117 (fixed) | 0.117(4) | 0.117 (fixed) |
| **L$_{micro}$ (μm)** | 2.06(5) | 2.09(5) | 2.1 (fixed) | 2.1 (fixed) |
| **D$_{micro}$ (×10$^{-4}$ Å$^{-3}$.cm$^{-1}$)** | 3.01(3) | 3.22(3) | 3.21(2) | 3.22(3) |
| **D$_{micro}$/L$_{micro}$ (×10$^{-8}$ Å$^{-4}$.cm$^{-1}$)** | 1.46(4) | 1.54(4) | 1.530(9) | 1.533(14) |
| **L$_{nano}$ (nm)** | 169.4(11) | 170.4(19) | 177.2(11) | 170.6(16) |
| **D$_{nano}$ (×10$^{-4}$ Å$^{-3}$.cm$^{-1}$)** | 13.12(8) | 14.39(11) | 13.70(8) | 14.40(11) |
| **D$_{nano}$/L$_{nano}$ (×10$^{-8}$ Å$^{-4}$.cm$^{-1}$)** | 77.4(7) | 84.4(11) | 77.3(7) | 84.4(10) |
| **χ$^2$ (%)** | 7.54 | 1.89 | 7.52 | 1.86 |

**Table S4**: Refined parameters and constraints for the bimodal DAB model using SANS and USANS data for lithium in the *in situ* cell after 2 cycles at 20 mA/cm$^2$. Estimated standard uncertainty for the last significant figures are given in parentheses.

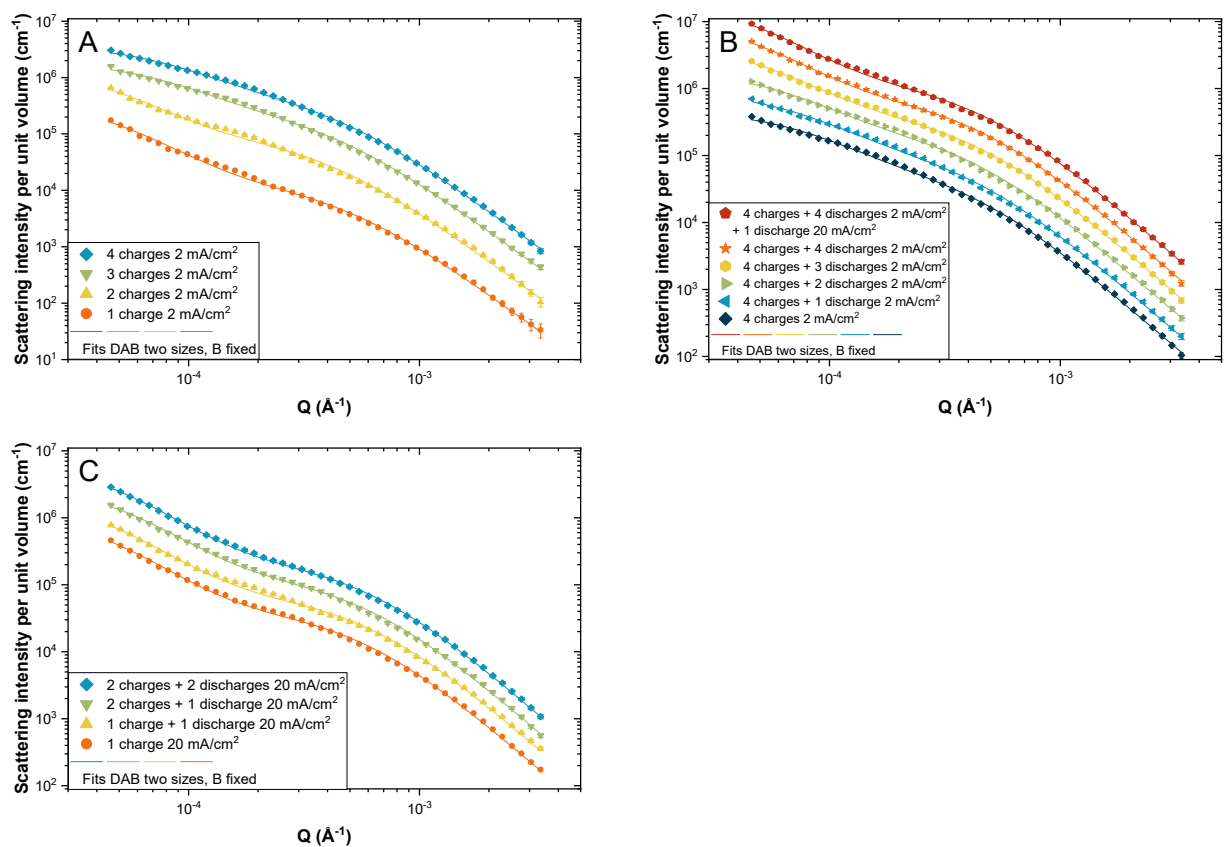

**Figure S7**: Slit-smeared USANS differential scattering cross-section per unit volume of lithium in the *in situ* cell after subtraction of scattering from electrochemically inactive components, shown offset in y for clarity where each pattern was multiplied by $1.8^{n-1}$ and n is the chronological pattern order in each graph starting at 1; the corresponding scattering data without offset were shown in Figure S5. Lines show the refined bimodal DAB model where B is fixed to 0.117 cm$^{-1}$. A) For the cell after one, two, three and four consecutive "charges" at 2 mA/cm$^2$, B) for the cell shown in A) after four "charges" and followed by one, two, three, and four consecutive "discharges" at 2 mA/cm$^2$, and an additional "discharge" at 20 mA/cm$^2$. C) For a cell after alternating "charges" and "discharges" at 20 mA/cm$^2$. Arrows are visual guides emphasizing intensity change.

| Dataset | $D_{micro}$ ($\times 10^{-4}$ Å$^{-3}$·cm$^{-4}$) | $L_{micro}$ (μm) | $D_{nano}$ ($\times 10^{-4}$ Å$^{-3}$·cm$^{-4}$) | $L_{nano}$ (nm) | $S_V$ 2 phases ($\times 10^3$ cm$^2$/cm$^3$) |
|---|---|---|---|---|---|
| 1 charge 2 mA/cm$^2$ | 1.05(2) | 2.17(12) | 2.94(4) | 218(5) | 3.43(8) |
| 2 charges 2 mA/cm$^2$ | 1.83(3) | 2.25(12) | 6.81(8) | 260(4) | 6.60(11) |
| 3 charges 2 mA/cm$^2$ | 3.50(17) | 1.22(5) | 10.56(19) | 246(6) | 11.2(2) |
| 4 charges 2 mA/cm$^2$ | 3.68(14) | 1.16(4) | 11.62(18) | 239(4) | 12.7(3) |
| 4 charges, 1 discharge 2 mA/cm$^2$ | 3.37(13) | 1.27(4) | 11.52(16) | 247(4) | 12.1(2) |
| 4 charges, 2 discharges 2 mA/cm$^2$ | 3.02(11) | 1.43(5) | 12.47(15) | 246(4) | 12.9(2) |
| 4 charges, 3 discharges 2 mA/cm$^2$ | 2.81(6) | 1.82(6) | 13.33(12) | 241(3) | 13.9(2) |
| 4 charges, 4 discharges 2 mA/cm$^2$ | 2.90(4) | 2.07(7) | 13.41(11) | 234(3) | 14.35(19) |
| 4 charges, 4 discharges 2 mA/cm$^2$ + 1 discharge 20 mA/cm$^2$ | 2.99(4) | 2.12(6) | 13.75(11) | 227(3) | 15.2(2) |
|  |  |  |  |  |  |
| 1 charge 20 mA/cm$^2$ | 2.78(3) | 2.30(7) | 13.63(11) | 180(2) | 18.8(2) |
| 1 charge, 1 discharge 20 mA/cm$^2$ | 2.69(3) | 2.23(7) | 14.33(11) | 171.0(19) | 20.8(3) |
| 2 charges, 1 discharge 20 mA/cm$^2$ | 3.20(4) | 1.94(5) | 14.47(11) | 173(2) | 20.9(3) |
| 2 charges, 2 discharges 20 mA/cm$^2$ | 3.22(3) | 2.09(5) | 14.39(11) | 170.4(19) | 21.1(3) |

**Table S5**: Refined parameters of the bimodal DAB model using USANS data for lithium in the *in situ* cell with B fixed to 0.117 cm$^{-1}$. The surface area per unit volume $S_V$ is calculated considering a single interface between lithium metal and electrolyte. Standard uncertainties estimated from the least-square regression analysis are shown in parentheses..

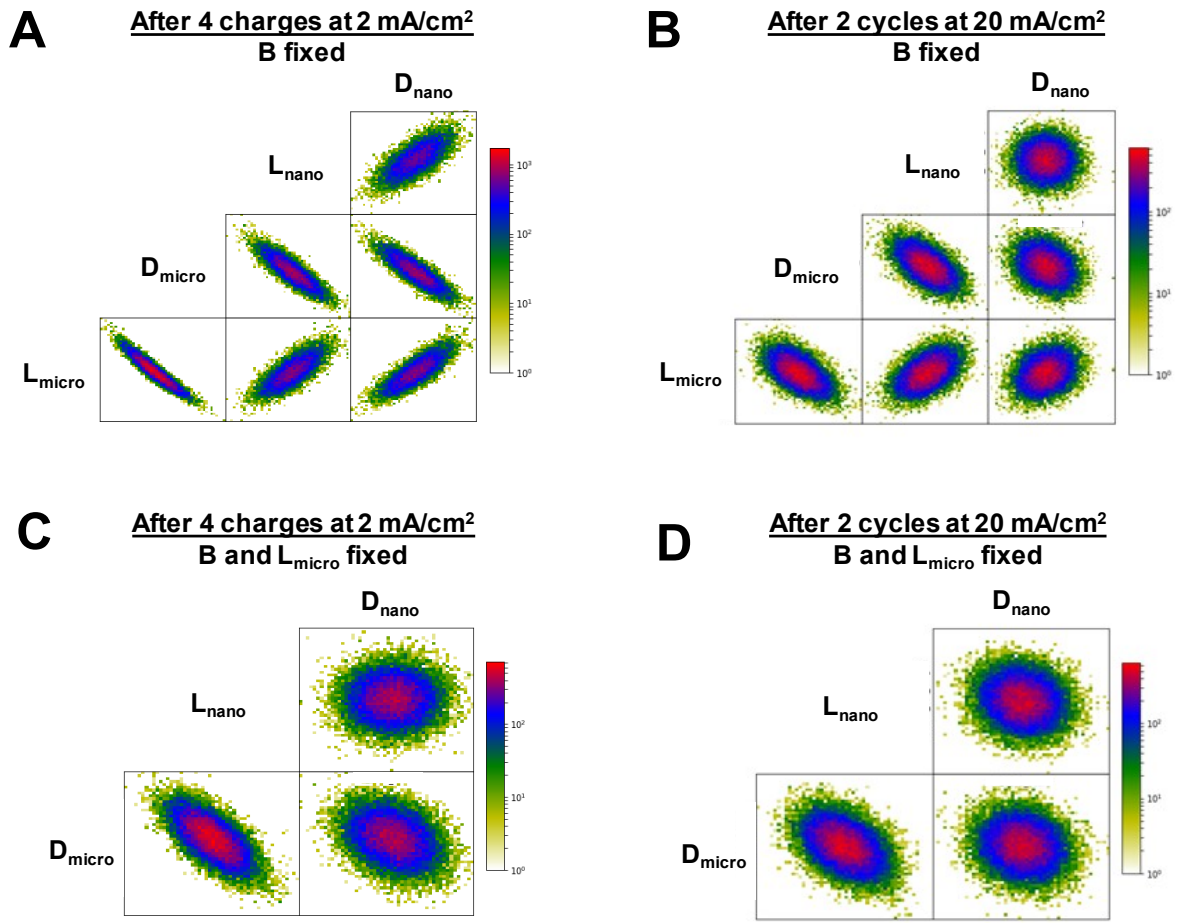

**Figure S8**: Correlation between refined parameters from the bimodal DAB model using USANS data for lithium in the *in situ* cell. A circular pattern indicates low correlation whereas a narrow line indicates strong correlation between fitting parameters. Correlations shown for models with A) B fixed to 0.117 cm$^{-1}$ for data after 4 charges at 2 mA/cm$^2$, with high correlations, B) B fixed to 0.117 cm$^{-1}$ after two cycles at 20 mA/cm$^2$ with moderate correlations, C) with B fixed to 0.117 cm$^{-1}$ and $L_{micro}$ fixed to 2.1 µm after 4 charges at 2 mA/cm$^2$, and D) with B fixed to 0.117 cm$^{-1}$ and $L_{micro}$ fixed to 2.1 µm after two cycles at 20 mA/cm$^2$.

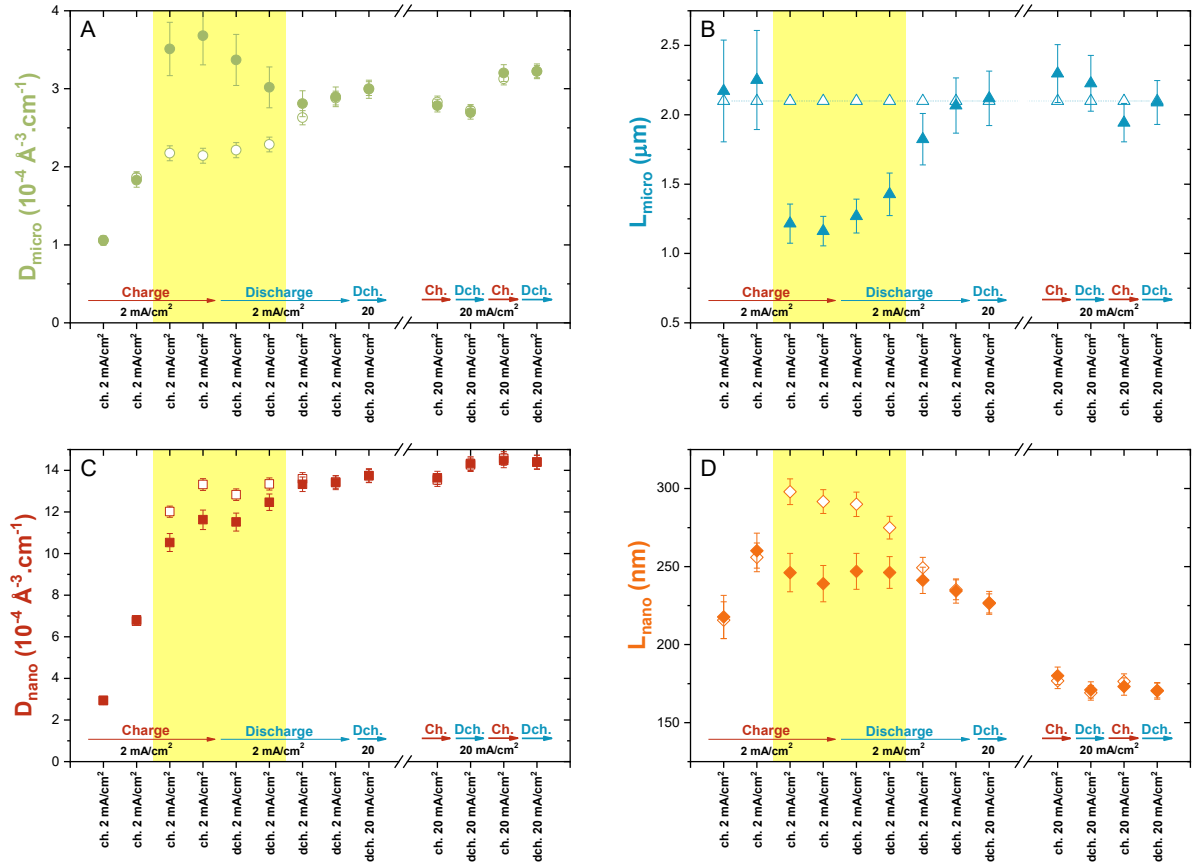

**Figure S9**: Refined parameters of the bimodal DAB model using USANS data for lithium in the *in situ* cell with B fixed to 0.117 cm$^{-1}$ (closed symbols) and B fixed to 0.117 cm$^{-1}$ and L$_{micro}$ fixed to 2.1 µm (open symbols). High parameter correlation was observed when L$_{micro}$ was refined using data after the 3$^{rd}$ charge, 4$^{th}$ charge, 1$^{st}$ discharge and 2$^{nd}$ discharge at 2 mA/cm$^2$, with highly correlated regions highlighted in yellow. Error bars are standard uncertainties estimated from the least-square regression analysis.

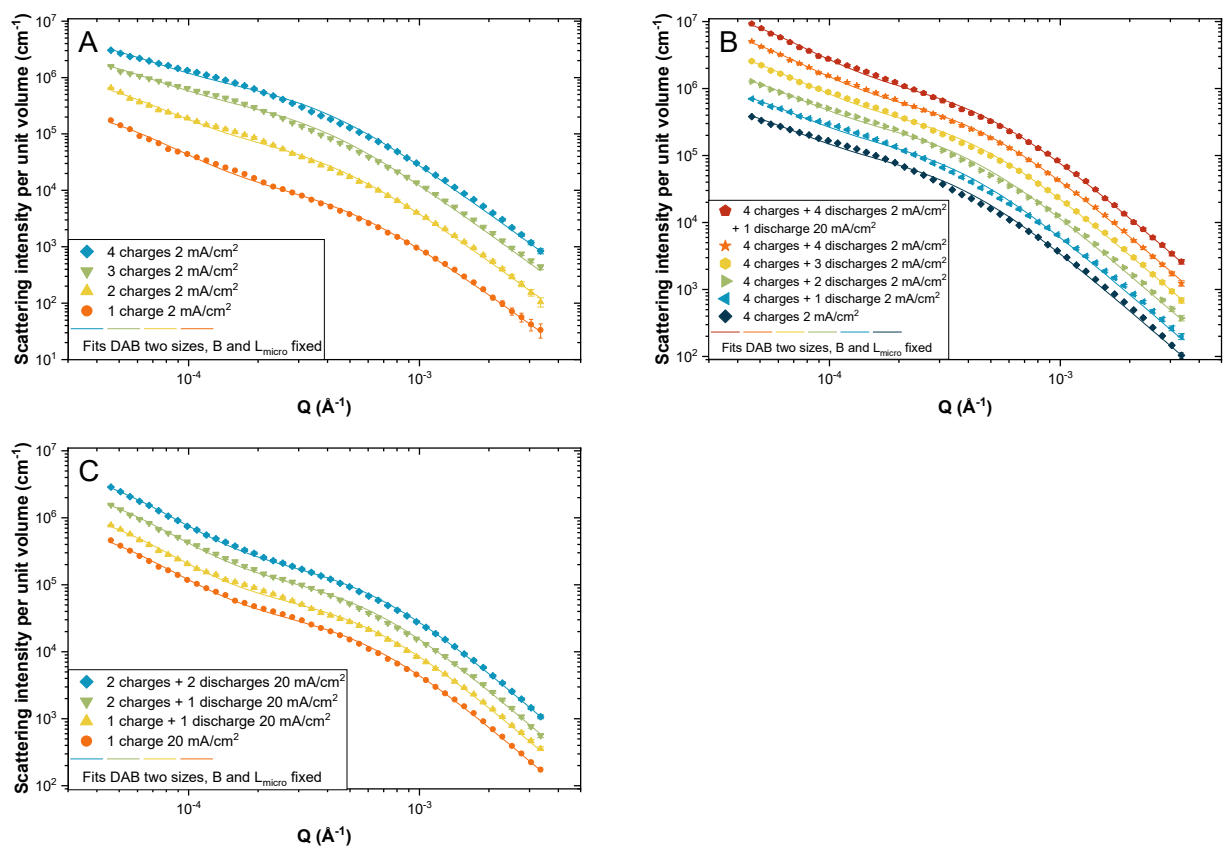

**Figure S10**: Slit-smeared USANS differential scattering cross-section per unit volume of lithium in the *in situ* cell after subtraction of scattering from electrochemically inactive components, shown offset in y for clarity, where each pattern was multiplied by $1.8^{n-1}$ and n is the chronological pattern order in each graph starting at 1; the corresponding scattering data without offset were shown in Figure S5. Lines show the refined bimodal DAB model with B fixed to $0.117\ cm^{-1}$ and $L_{micro}$ to 2.1 μm using data A) for the cell after one, two, three and four consecutive "charges" at $2\ mA/cm^2$, B) for the cell shown in A) after four "charges" and followed by one, two, three and four consecutive "discharges" at $2\ mA/cm^2$, and an additional "discharge" at $20\ mA/cm^2$, C) for a cell after alternating "charges" and "discharges" at $20\ mA/cm^2$. Arrows are visual guides emphasizing intensity change.

| Dataset | $D_{micro}$ ($\times 10^{-4}$ Å$^{-3}$·cm$^{-4}$) | $D_{nano}$ ($\times 10^{-4}$ Å$^{-3}$·cm$^{-4}$) | $L_{nano}$ (nm) | $S_V$ 2 phases ($\times 10^3$ cm$^2$/cm$^3$) |
|---|---|---|---|---|
| 1 charge 2 mA/cm$^2$ | 1.060(18) | 2.93(4) | 216(4) | 3.45(7) |
| 2 charges 2 mA/cm$^2$ | 1.86(2) | 6.74(6) | 256(3) | 6.67(10) |
| 3 charges 2 mA/cm$^2$ | 2.17(3) | 12.01(9) | 298(3) | 10.12(12) |
| 4 charges 2 mA/cm$^2$ | 2.14(3) | 13.32(10) | 292(3) | 11.42(13) |
| 4 charges, 1 discharge 2 mA/cm$^2$ | 2.21(3) | 12.83(9) | 290(3) | 11.09(13) |
| 4 charges, 2 discharges 2 mA/cm$^2$ | 2.29(3) | 13.35(10) | 275(2) | 12.15(14) |
| 4 charges, 3 discharges 2 mA/cm$^2$ | 2.63(3) | 13.59(10) | 249(2) | 13.65(15) |
| 4 charges, 4 discharges 2 mA/cm$^2$ | 2.88(3) | 13.44(10) | 235(2) | 14.32(16) |
| 4 charges, 4 discharges 2 mA/cm$^2$ + 1 discharge 20 mA/cm$^2$ | 3.00(3) | 13.73(10) | 226(2) | 15.20(17) |
| | | | | |
| 1 charge 20 mA/cm$^2$ | 2.83(3) | 13.54(10) | 176.8(13) | 19.1(2) |
| 1 charge, 1 discharge 20 mA/cm$^2$ | 2.72(2) | 14.27(11) | 169.1(16) | 21.0(2) |
| 2 charges, 1 discharge 20 mA/cm$^2$ | 3.13(3) | 14.58(11) | 176.4(16) | 20.6(2) |
| 2 charges, 2 discharges 20 mA/cm$^2$ | 3.22(3) | 14.40(11) | 170.6(16) | 21.0(2) |

**Table S6**: Refined parameters of the bimodal DAB model using USANS data for lithium in the *in situ* cell with B fixed to 0.117 cm$^{-1}$ and $L_{micro}$ fixed to 2.1 μm. The surface area per unit volume $S_V$ is calculated considering a single interface between lithium metal and electrolyte. Standard uncertainties estimated from the least-square regression analysis are shown in parentheses.

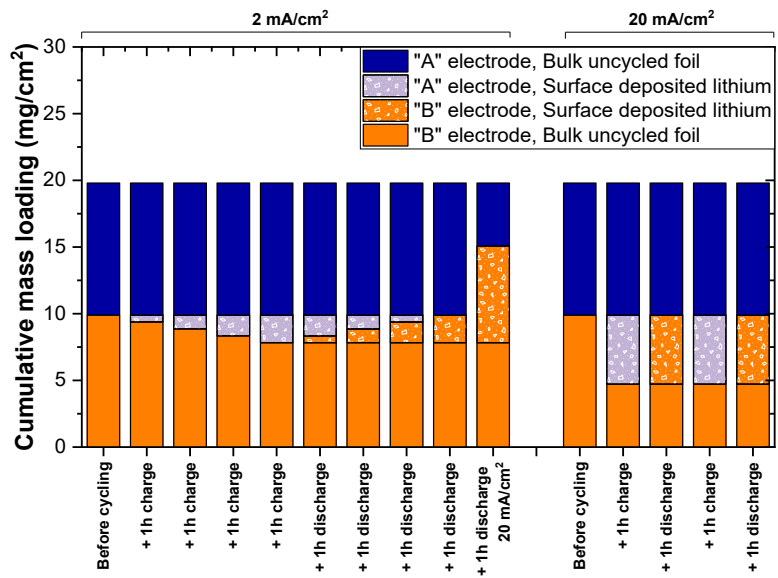

**Figure S11**: Cumulative mass loading of excess (uncycled) bulk lithium foil and of deposited lithium on each side of the battery following electrochemical cycling, assuming no loss of capacity and complete resorption of previously deposited lithium. The "A" and "B" electrodes are arbitrary in the symmetrical lithium cell with the term "A" identifying the electrode where lithium was first deposited.

**Technical Note S3**: Calculation of the scattering for lithium in the *in situ* cell by subtraction of scattering measured for each other component.

The differential scattering cross-section per unit volume for the *in situ* cell corresponds to the sum of scattering of each component multiplied by their thickness:

$$I_V(in\ situ\ cell) \times [t(Li) + t(Cu) + t(PVDF) + t(pouch)]$$
$$= I_V(Li) \times t(Li) + I_V(Cu) \times t(Cu) + I_V(PVDF) \times t(PVDF) + I_V(pouch) \times t(pouch)$$

Where Li, Cu, PVDF and pouch denote lithium foils, smooth copper foils, electrolyte-wet PVDF and the laminated pouch. Scattering per unit volume for lithium in the *in situ* cell is obtained by rearranging the equation:

$$I_V(Li) = \begin{bmatrix} I_V(in\ situ\ cell) \times [t(Li) + t(Cu) + t(PVDF) + t(pouch)] \\ -I_V(Cu) \times t(Cu) - I_V(PVDF) \times t(PVDF) - I_V(pouch) \times t(pouch) \end{bmatrix} \div t(Li)$$

Using thicknesses as given in §7.1 of the Methods section, with $t(Li) = 2\times200$ μm, $t(Cu) = 2\times25$ μm, $t(PVDF) = 110$ μm and $t(pouch) = 2\times115$ μm.